\documentclass[apj]{emulateapj}
\usepackage{amssymb,amsmath}
\usepackage{epsfig}

\shorttitle{HOT HALO OF NGC\,891}
\shortauthors{HODGES-KLUCK \& BREGMAN}

\begin{document}
\title{A Deep X-ray View of the Hot Halo in the Edge-on Spiral Galaxy NGC\,891}

\author{Edmund~J.~Hodges-Kluck$^{1}$ \& Joel~N.~Bregman$^{1}$}

\altaffiltext{1}{Department of Astronomy, University of Michigan, Ann
  Arbor, MI 48109}
\email{hodgeskl@umich.edu}

\begin{abstract}
NGC\,891 is a nearby edge-on galaxy that is similar to the Milky Way and
has a hot X-ray emitting halo that could arise from accretion, a galactic
fountain, or a combination of the two. The metallicity of the gas can help 
distinguish between these models, and here we report on results that use 
138\,ks of archival \textit{Chandra} data and 92\,ks of new \textit{XMM-Newton} 
data to measure the temperature and metallicity of the hot halo of the galaxy.  We
find good fits for a thermal model with $kT \sim 0.2$\,keV and $Z \sim
0.1 Z_{\odot}$, and rule out solar metallicity to more than 99\%
confidence.  This result suggests accretion from the intergalactic medium 
as the origin for the hot halo.  However, it is also
possible to fit a two-temperature thermal model with solar metallicity
where $kT_1 \sim 0.1$\,keV and $kT_2 \sim 0.25$\,keV.  A consideration of the 
cooling rate and scale height prefers the single-temperature model.
We also find that the cooling rate in the hot gas cannot explain the
massive \ion{H}{1} halo in the steady state.  In addition, a
galactic fountain model cannot eject enough mass to account for
the \ion{H}{1} halo, and we speculate that the neutral halo may be gas 
from a prior outflow that has since cooled.
\end{abstract}


\section{Introduction}
\label{section.introduction}

The hot halos of spiral galaxies like our own are reservoirs of
$1-3\times 10^6$\,K gas whose origin is unclear.  These halos are
potentially important to galaxy formation theories because some 
closed-box models of Galactic chemical evolution predict a larger
number of metal-poor stars in the solar neighborhood than we 
observe \citep[the G-dwarf or M-dwarf problem, see][]{vandenbergh62,woolf12},
and this issue is not isolated to the Milky Way \citep[e.g.][]{worthey96}.
This problem can be solved by the accretion of low-metallicity gas
onto the disk from the halo.  
In addition, \citet{sancisi08} argue that about $1 M_{\odot}$\,yr$^{-1}$
of accretion onto the disk is required to sustain the star formation rate 
(SFR) of typical spirals; \citet{binney00} find that the SFR in the solar
neighborhood has been nearly constant for the past 11\,Gyr for a Salpeter
initial mass function, so star formation may indeed be sustained for long
times.  It is not yet clear whether the required accretion occurs primarily
via condensation of a radiatively cooling hot halo or cold flows
and minor mergers \citep[for a review of cold accretion, see][]{sancisi08}.

It is also unclear whether the hot halo is gas accreted from the local
intergalactic medium (IGM) as suggested by \citet{cox76}.  Instead,
they may be gas expelled from the disk by supernovae (SNe) in a
``galactic fountain'' \citep{shapiro76,bregman80}.  Both can explain
the observed hot halos: in the case of accretion, infalling gas from
the IGM shocks and is heated to the dynamical temperature, whereas a
galactic fountain forms the hot halo as superbubbles powered by
hundreds of SNe break out of the galactic disk and vent their contents
to the halo.

Since both the galactic fountain and accretion from the IGM are
expected to occur, it is not clear which contributes most to the hot
halo. For example, if the supernova rate is too low, the bubbles may
never break out or only contribute to a halo with a small scale
height.  As both mechanisms can produce a halo with a similar
temperature and disk condensation rate, the simplest way to
distinguish them is the metallicity of the hot gas, which should be
close to the IGM value in the accretion scenario \citep[$\sim 0.05-0.3
Z_{\odot}$ in a spiral-rich cluster such as HCG\,16;][]{belsole03},
and solar or super-solar in the fountain scenario.

The hot halos are visible in the soft X-rays where the metallicity can
be measured directly.  Since edge-on spirals allow for clean separation of
the disk and halo emission while amplifying the surface brightness of the
halo, they are the ideal candidates for this measurement.

In this paper, we present an X-ray study of the hot halo of NGC\,891
using \textit{Chandra} and \textit{XMM-Newton} data to measure the
halo metallicity.  NGC\,891 is a nearby \citep[$d \sim
10$\,Mpc;][]{temple05} edge-on spiral thought to be a close Milky Way
analog in luminosity and spectral type \citep[e.g.][]{vanderkruit81},
and, as in our Galaxy, there is dynamical evidence for a bar
\citep{garcia-burillo95}.  NGC\,891 is not interacting with other
galaxies, is almost perfectly edge-on, and has no optical or \ion{H}{1} warp
\citep{rupen91,swaters97}, allowing an unambiguous distinction between
halo and disk emission.  However, it is known to be slightly lopsided
\citep{sancisi79}.  The primary aim of our study is
to measure the metallicity of the hot halo directly from the X-ray
data.  A summary of basic parameters for NGC\,891 is given in
Table~\ref{table.basic}.  

\begin{figure*}
\begin{center}
\includegraphics[width=1.0\linewidth]{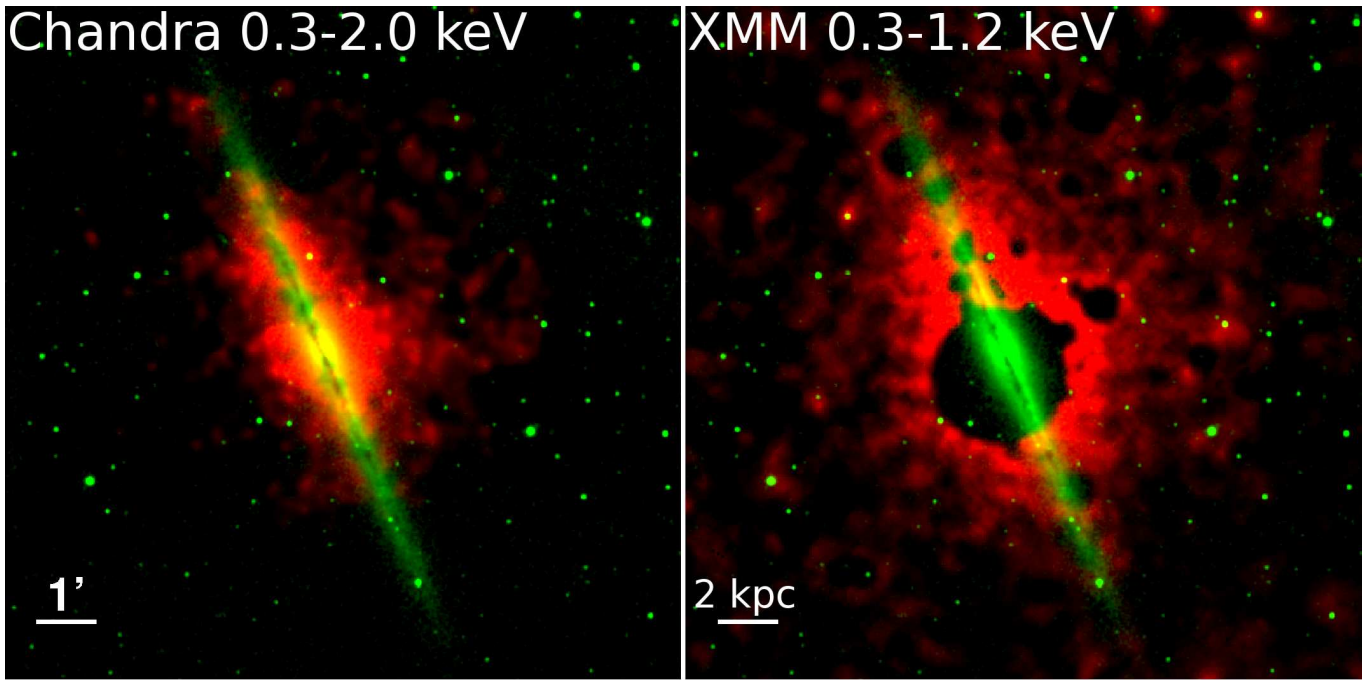}
\caption{Left: Composite 2MASS $H$-band image of NGC\,891 (green) with
bright emission from the adaptively smoothed $0.3-2.0$\,keV \textit{Chandra}
image (red).  Right: Same as left, but with $0.3-1.2$\,keV combined 
adaptively smoothed \textit{XMM-Newton} image in red instead.  
Te dark holes are point source masks shown in
Figure~\ref{figure.xmm_masks} (the large one is the ULX).}
\label{figure.inner_outer_halo}
\end{center}
\end{figure*}

\begin{figure*}
\begin{center}
\includegraphics[width=1.0\linewidth]{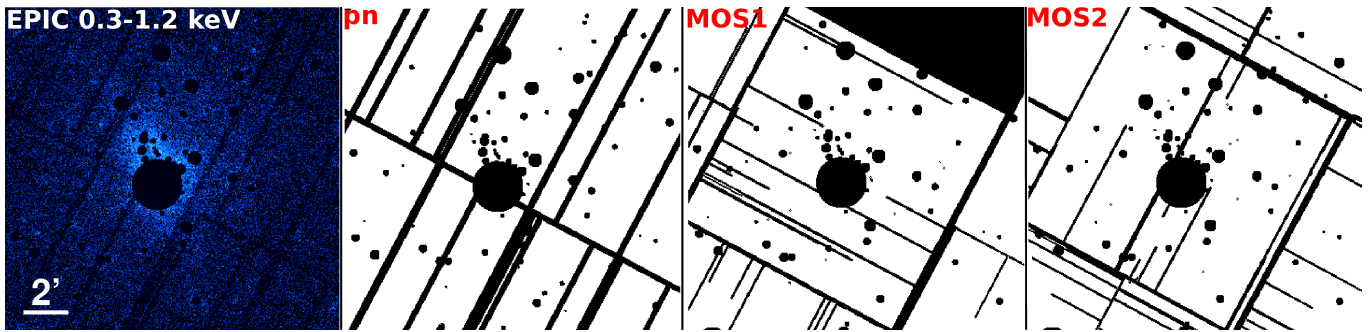}
\caption{At left we show the combined EPIC $0.3-1.2$\,keV image with
  point sources removed before exposure correction.  The remaining
  three panels show the Swiss-cheese point-source and chip-gap/bad pixel masks
  created with XMM-ESAS, centered on the galaxy.  The
  \textit{XMM-Newton} image in Figure~\ref{figure.inner_outer_halo} is
  a smoothed, exposure corrected version of the counts image shown
  here.}
\label{figure.xmm_masks}
\end{center}
\end{figure*}

NGC\,891 is of particular interest because it is the X-ray brightest of 
normal edge-on galaxies \citep{bregman97} and has been held up as a prime
example of a galactic fountain in action.  The evidence in favor of such
a fountain is quite strong, enabling us to determine whether the X-ray
emitting gas is an important participant in the fountain or largely 
decoupled accreted material that slowly settles onto the disk.  We review
the evidence presently.

First, the disk appears to be dynamically connected to the halo.  The
cold \ion{H}{1} halo rotates differentially, with rotation velocity decreasing
with distance from the disk \citep{oosterloo07}, and the spatial
structure of the radio emission correlates with that of the H$\alpha$
halo \citep{dettmar90} and the X-rays
\citep{bregman94,strickland04}, being concentrated in the north and
near the disk.  The SFR is also highest here, which \citet{tullmann06} argue
leads naturally to the observed distribution since the amount of hot
gas is proportional to the energy injected by supernovae.  This is
consistent with the dust halo \citep{howk97,howk00}, which shows more
pronounced filaments on the northern side (although perhaps just as
many on the southern side), but we note that \citet{garcia-burillo92}
find that the extended molecular halo is consistent with being
symmetric about the rotation axis.  

Second, there is strong evidence for magnetic fields lacing
the halo that are naturally explained by a galactic fountain.  There
is radio synchrotron emission cospatial with the \ion{H}{1} halo out to
10\,kpc away from the disk \citep{oosterloo07}, and there are
collimated H$\alpha$ filaments that extend a few kpc above the disk
\citep{rossa04}.  This suggests that the magnetic field originates in
the disk \citep{dahlem94}.  

Finally, clear evidence of supernova activity is visible in the disk.
\citet{rossa04} find superbubbles and shells near the midplane, as
well as filamentary structures that appear to be connected to
the disk superbubbles.  These may signify breakouts from the disk
where the hot medium can rise into the halo.

The rest of this paper is organized as follows.  In
Section~\ref{section.obs}, we describe the X-ray observations of
NGC\,891 and how we processed the resultant data.
Section~\ref{section.metallicity} contains the temperatures and
metallicities we derive from fits to the spectra, as well as estimates
of the uncertainties in these parameters from various systematic
effects.  We present two competing models that cannot be
spectroscopically distinguished with the present data, so in the first
part of Section~\ref{section.discussion} we appeal to the expected
cooling rate and morphology of the hot halo in each model.  Afterward,
we consider more generally whether the hot gas is coupled to the
neutral halo and whether the system is in a steady state.  We conclude
by summarizing our results and arguments.

\section{X-ray Observations \& Processing}
\label{section.obs}

We processed new and archival X-ray data from the \textit{Chandra}
X-ray Observatory and \textit{XMM-Newton}
(Table~\ref{table.observations}).  Our primary goal was to measure the
metallicity of the hot halo, which is visible in the soft X-rays.  To
do so, we extract spectra from regions on the detectors corresponding
to the halo and fit them with a thermal plasma model, assuming
collisional ionization equilibrium.  In the remainder of this section,
we describe the observations and processing, surface brightness maps,
and our spectral fitting method.


\begin{deluxetable}{lll}
\tablenum{1}
\tabletypesize{\scriptsize}
\tablecaption{Basic Parameters for NGC\,891}
\tablewidth{0pt}
\tablehead{
\colhead{Parameter} & \colhead{Units} & \colhead{Value}
}
\startdata
$\alpha$ (J2000)\tablenotemark{a}  & & 02h22m33.4s\\
$\delta$ (J2000)\tablenotemark{a}  & & +42d20m57s\\
$d$\tablenotemark{b}               & (Mpc) & 9.8\\
$z$\tablenotemark{c}               & & 0.00176\\
$i$ \tablenotemark{d}              & (deg) & 89\\
Scale\tablenotemark{a}             & (kpc arcmin$^{-1}$) & 2.2\\
Morphology\tablenotemark{e}        & & SBb\\
$M_{\text{HI}}$\tablenotemark{d}   & ($M_{\odot}$) & $1.2\times 10^9$\\
$M_{\text{tot}}$\tablenotemark{d}  & ($M_{\odot}$) & $1.4\times 10^{11}$\\
$L_X$\tablenotemark{f}             & (erg s$^{-1}$)& $4.4\times 10^{39}$\\
$m_V$\tablenotemark{a}             & (mag) & 10.81\\
SFR\tablenotemark{g}               & ($M_{\odot}$\,yr$^{-1}$) & 3.8
\enddata
\tablerefs{\label{table.basic} (a) NED (b) \citet{tikhonov05} (c) \citet{devaucouleurs91}
  (d) \citet{oosterloo07} (e) \citet{garcia-burillo95} (f) \citet{bregman94}
  (g) \citet{popescu04} }
\end{deluxetable}

\subsection{\textit{Chandra} Reduction}

We used two archival \textit{Chandra} observations from 2000-11-01
(obsID 794) and 2003-12-10 (obsID 4613) with a combined exposure time
of 170\,ks.  These data have previously been published by
\citet{strickland04} and \citet{temple05}.  The data were reprocessed
using standard \textit{Chandra} Interactive Analysis of Observations
(CIAO v4.4) software recipes to create level=2 files.  We extracted
light curves from background regions in the $0.3-10$~keV range to look
for flaring exceeding 3$\sigma$ from the mean count rate.  A large
flare affected the last 23\,ks of obsID 794, and intermittent short
flares affect about 5\,ks of obsID 4613, leaving 138\,ks of good time.
Our analysis is confined to the exposure-corrected ACIS-S3 chip except
where otherwise mentioned.  We then used the {\sc wavdetect} algorithm
implemented in CIAO to create point source masks in the $0.3-8$\,keV
bandpass, with manual filtering for anomalous regions near the chip
edges.  In practice, this eliminates point sources down to a flux of
$F_X \sim 5\times 10^{-16}$\,erg\,s$^{-1}$\,cm$^{-2}$ in the
$0.3-8$\,keV band near the center of the merged exposure (where the
PSF is small) based on the faintest sources detected there.  The 5$\sigma$ 
point source sensitivity in the $0.3-8$\,keV band for the
merged exposures is about $F_X \approx 3.8\times 10^{-16}$\,erg\,s\,cm$^{-2}$. 


\begin{deluxetable*}{llllll}
\tablenum{2}
\tabletypesize{\scriptsize}
\tablecaption{X-ray Observations of NGC\,891 in this paper}
\tablewidth{0pt}
\tablehead{
\colhead{Date} & \colhead{Observatory} & \colhead{Detectors} & \colhead{Exp. Time} &
\colhead{Sum GTIs} & \colhead{Obs. ID}  \\
& & & & \colhead{(ks)} & \colhead{(ks)}
}
\startdata
1993-08-27 & ROSAT      & PSPC        & 31    & 31  & RP500266N00\\
2000-11-01 & Chandra    & ACIS-S      & 51    & 30  & 794        \\
2003-12-10 & Chandra    & ACIS-S      & 121   & 108 & 4613       \\
2011-08-25 & XMM-Newton & EPIC MOS/pn & 133   & 92  & 0670950101 
\enddata
\label{table.observations}
\end{deluxetable*}

\begin{figure}
\begin{center}
\includegraphics[width=1.0\linewidth]{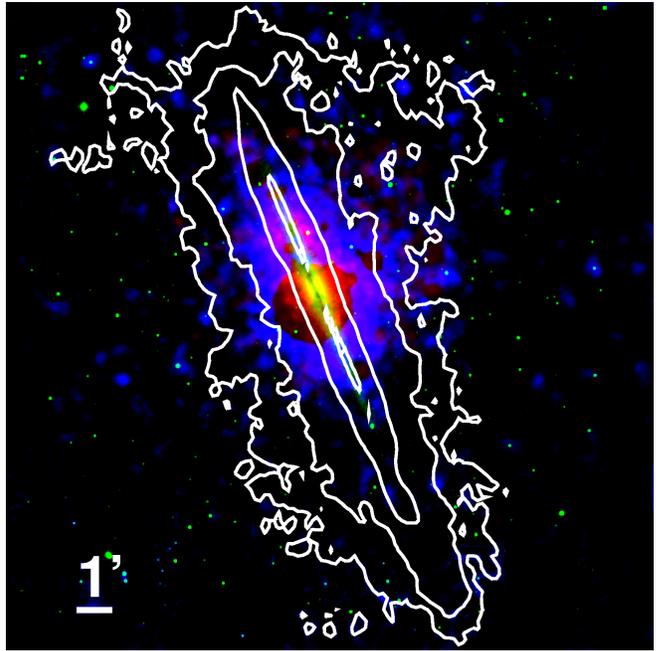}
\caption{Red: Inner halo from $0.3-2.0$\,keV adaptively smoothed \textit{Chandra} image.  
Green: 2MASS $H$-band image.  Blue: $0.3-1.2$\,keV adaptively smoothed 
\textit{XMM-Newton} image showing the outer halo.  Contours: Total \ion{H}{1} column density, where contours are
$10^{19}$, $10^{20}$, $10^{21}$, and $10^{22}$\,cm$^{-2}$.  The Galactic
column towards the galaxy is approximately $6.5\times 10^{20}$\,cm$^{-2}$. }
\label{figure.HI_halo}
\end{center}
\end{figure}

\begin{figure}
\begin{center}
\includegraphics[width=0.8\linewidth]{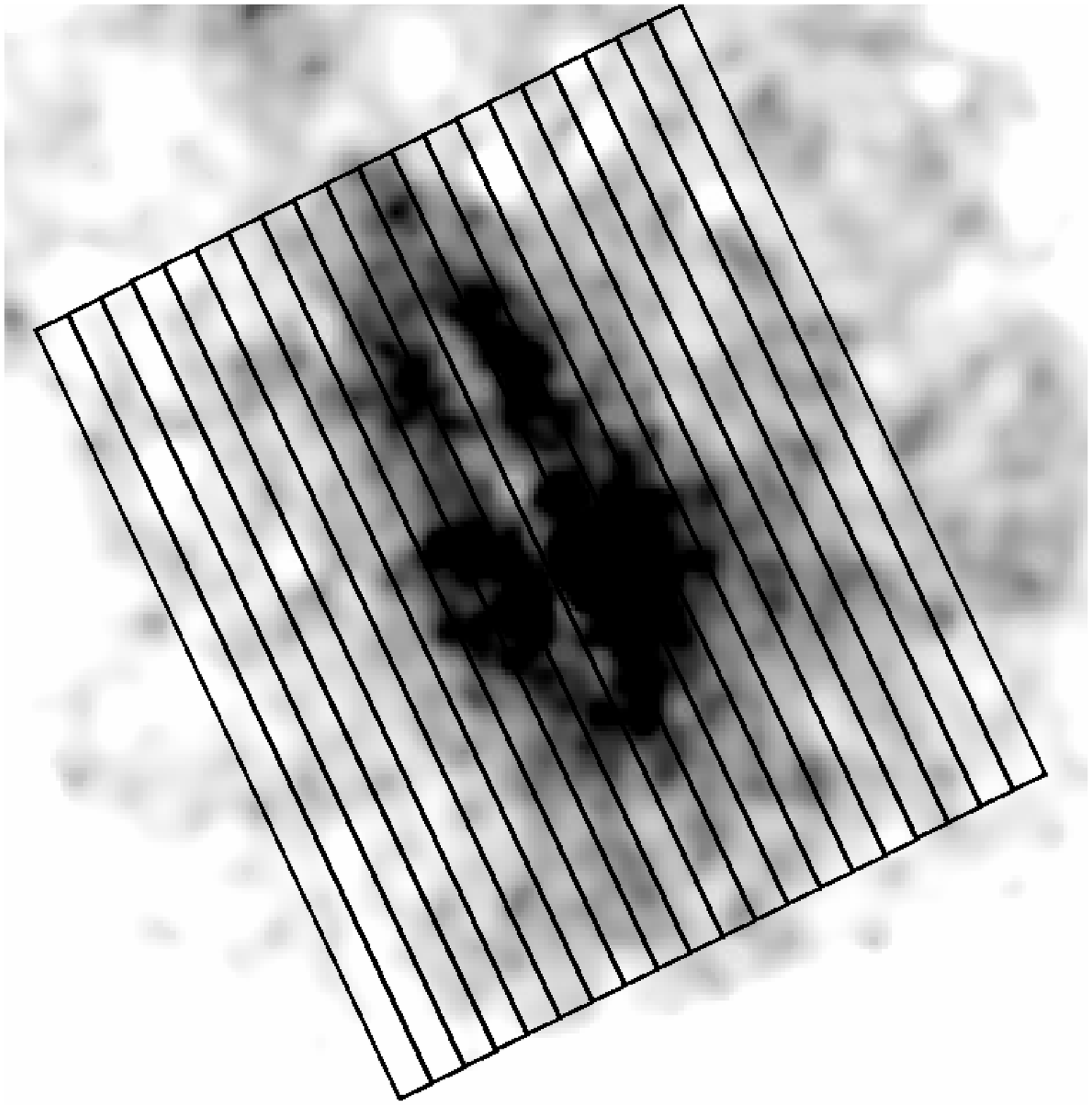}
\includegraphics[width=0.75\linewidth,angle=90]{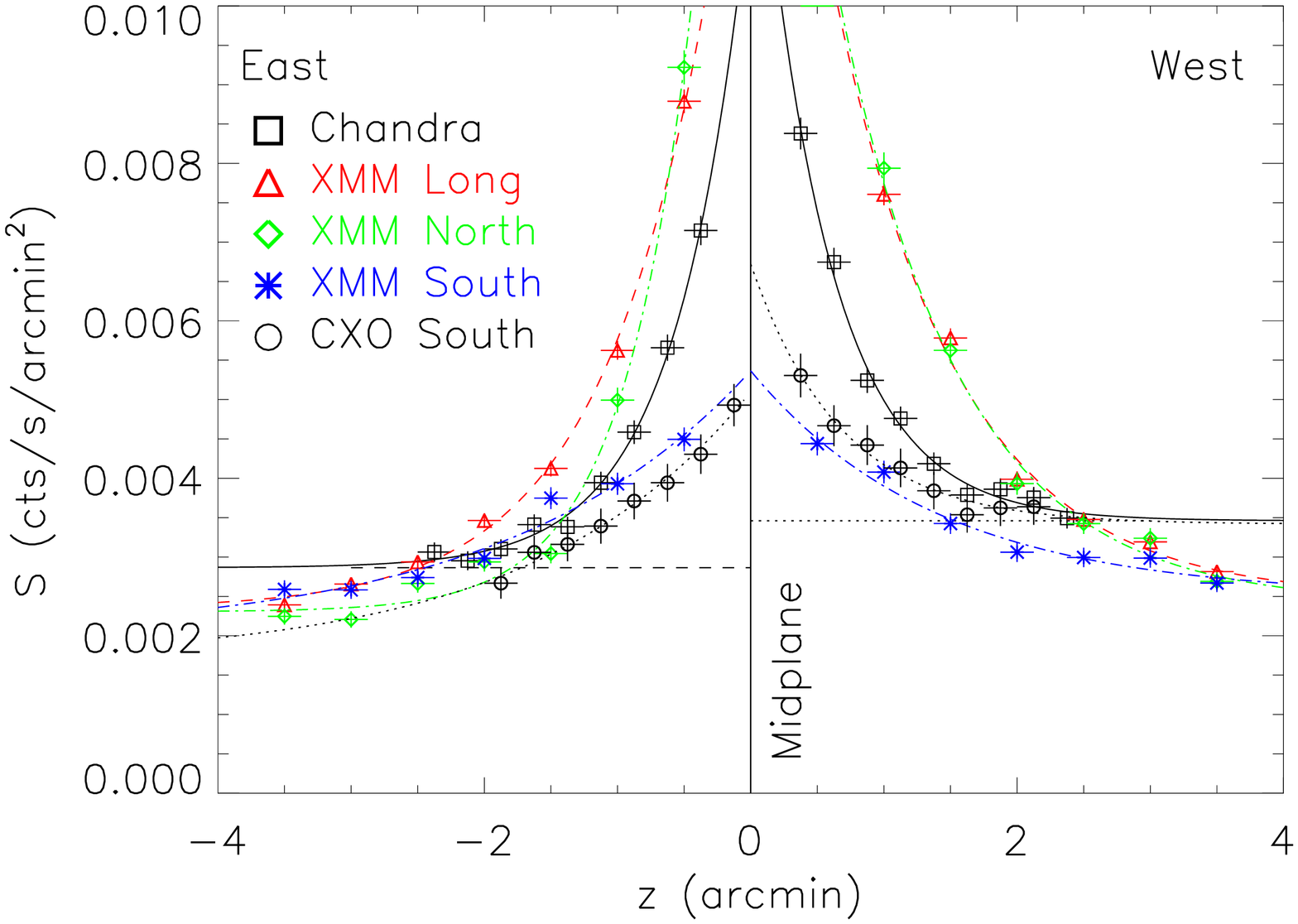}
\caption{The bottom plot shows surface brightness as a function of height $z$ above the
  galactic midplane in the \textit{Chandra} $0.3-2.0$\,keV data (black
  squares) and \textit{XMM-Newton} $0.3-1.2$\,keV data.  The
  adaptively smoothed \textit{Chandra} image is shown with the
  extraction boxes overlaid.  Point sources have been subtracted, but
  all diffuse emission is included, and the values are blindly
  corrected for the point source masks by scaling the true enclosed
  area to the box area.  Exponential fits are overlaid.  See text for discussion.}
\label{figure.midplane_profile}
\end{center}
\end{figure}

\subsection{\textit{XMM-Newton} Reduction}

We obtained a 133\,ks \textit{XMM-Newton} observation of NGC\,891 on
2011-08-25 (obsID 0670950101).  An archival shorter exposure has been
published by \citet{temple05}, but it is severely affected by flaring
and shows diffuse emission only near the bulge, so we omit it from
this study.  The data were obtained using the ``medium'' filter, meaning
at 1\,keV the total EPIC effective area is about 1200\,cm$^{-2}$ (cf.
555\,cm$^{-2}$ for \textit{Chandra}).
The data were processed using the standard reduction
tasks for the EPIC cameras in the Scientific Analysis System software
(SAS v11.0) and screened for proton flaring, after which 92\,ks of
good time remains.  Unlike the \textit{Chandra} images, this exposure
is severely contaminated by the presence of a bright new source with
an absorbed $0.3-10$\,keV flux of $F_X \approx 1.6\times
10^{-12}$\,erg\,s$^{-1}$ near the disk of the galaxy.  This source,
which we argue is an ultraluminous X-ray source (ULX) in NGC\,891
\citep{hodges-kluck12}, was not present in the \textit{Chandra}
observations nor any archival X-ray observation publicly available.

After standard processing, including exposure correction, we created Swiss-cheese 
point-source masks using the \textit{cheese} and \textit{cheese-bands}
scripts in the \textit{XMM-Newton} Extended Source Analysis Software\footnote{see
http://heasarc.gsfc.nasa.gov/docs/xmm/xmmhp\_xmmesas.html} (XMM-ESAS), which are
included in SAS \citep{snowden04}.  We used a cutoff flux of $F_X =
5\times10^{-16}$\,erg\,s$^{-1}$\,cm$^{-2}$ in the $0.4-7.2$\,keV
bandpass.  The \textit{XMM-Newton} point source sensitivity for this
exposure is about $4\times 10^{-16}$\,erg\,s$^{-1}$\,cm$^{-2}$ in the pn and
slightly worse in the MOS detectors\footnote{for nominal values, see
http://xmm.esa.int/external/xmm\_user\_support/documentation/uhb\_2.5/node38.html}.

When compared to the masks produced for the
\textit{Chandra} data, we find good agreement near the central region
except for the ULX, which we excise to the 90\% encircled energy
radius (1\,arcmin).  We also note that chip 5 on the MOS\,1 detector
has anomalously high background and is removed from analysis; however,
no significant source emission falls on this chip.

\subsection{X-ray Images \& Scale Height}

The diffuse emission associated with the hot halo can be seen in
adaptively smoothed maps of the soft emission with point sources
removed (Figure~\ref{figure.inner_outer_halo}; the large cavity on the
right is the ULX mask, and the point source masks for each 
EPIC camera are shown in Figure~\ref{figure.xmm_masks}).  The left panel shows
the bright emission close to the disk seen by \textit{Chandra}.
Little emission comes from the disk itself (where the soft X-rays from
the ISM are absorbed by the high $N_H$ column), and the emission
immediately outside is dominated by the bulge near the galactic
center.  The non-bulge emission is primarily seen to the north of the
galactic center and is concentrated within about 1.5\,kpc of the disk.
This is the halo emission most obvious at lower sensitivities
\citep{bregman94}, and it correlates with the diffuse ionized
(H$\alpha$) halo \citep{dettmar90} and higher SFR on the northern
side.  In the deeper \textit{XMM-Newton} image to the right, it is
evident that the diffuse X-ray emission extends much farther from the
bulge \citep[also described in the \textit{ROSAT} observation
in][]{bregman94} and is more symmetric on larger scales.  The
diffuse emission is much more evident in the \textit{XMM-Newton}
image in Figure~\ref{figure.inner_outer_halo} due to the higher
effective area of the telescope and the moderate angular resolution
(6\,arcsec) compared to \textit{Chandra} (0.5\,arcsec).  

The difference between the emission on large and small scales means
that it is possible that the physical properties of the emitting gas
are also different.  For fitting purposes it is also convenient to
extract spectra from regions near the disk (where the \ion{H}{1}
absorption is strong and variable, see Figure~\ref{figure.HI_halo})
and regions where the dominant $N_H$ column is the Galactic one.
Since $N_H$ is somewhat degenerate with $kT$ and $Z$ at CCD spectral
resolution, the cleanest measurement of $Z$ can be made in the latter
case.  We therefore divide the X-ray emission into two regions: the
``inner halo'' (where absorption must be modeled explicitly) and the
``outer halo'' (where we use the Galactic column only).  The inner
halo is defined as the region more than two optical scale heights
above the disk midplane \citep[$H \sim 200-400$\,pc in][and we adopt
300\,pc]{xilouris98} but below where the \ion{H}{1} absorbing column
falls to half the Galactic value of $N_H \approx 6.5\times
10^{20}$\,cm$^{-2}$ \citep{kalberla05}, which is between 1.5--3\,kpc
depending on position along the disk (Figure~\ref{figure.HI_halo}).
The inner boundary is chosen to exclude X-ray binary emission, as ISM
disk emission is negligible.  The outer halo is simply the region
above this (where the Galactic column dominates absorption) out to
where the source emission falls below the background.

In practice, defining the rotation axis of the galaxy as $\hat{z}$ and
the projected midplane axis as $\hat{r}$, the inner halo has
boundaries $0.3 \text{ arcmin}<z<1.3 \text{ arcmin}$
(0.6\,kpc$<z<$2.8\,kpc) and the outer halo $z>1.3\text{ arcmin}$
($z>2.8$\,kpc).  For extraction purposes, we make boxes whose length
in $\hat{r}$ is the projected disk length (7\,arcmin or
$\sim$15\,kpc).  

Filamentary structure and signs of outflows are common in X-ray images
of edge-on (mostly starburst) galaxies \citep[e.g.][]{strickland04}.
We searched for structure by making unsharp masks of the
\textit{Chandra} and \textit{XMM-Newton} data and looking for overlap.
Although apparent filaments are visible in each dataset, there is
little agreement between images.  On closer inspection, the putative
filaments often include emission near point source masks.  Because the
PSF wings produce enhanced surface brightness over the background
while the mask itself is a decrement, filamentary structure can be
mimicked when a subtracted source is near two (intrinsically
uncorrelated) regions with higher surface brightness.  We also find
that several of the structures in the \textit{XMM-Newton} map include
point sources that were too faint or distorted to be identified by the
\textit{cheese} algorithm but are obvious in the \textit{Chandra}
data.  We find no evidence for outflow-like or filamentary structures
in the data, but as the emission is integrated through the halo we
cannot rule it out.

The outer halo is approximately symmetric about $\hat{z}$ and
$\hat{r}$ and is best seen in the \textit{XMM-Newton} image
(Figure~\ref{figure.inner_outer_halo}), as the outer halo is about as
wide as the \textit{Chandra} ACIS-S3 chip.  To estimate the scale
height, we use the exposure-corrected, point-source subtracted data in
the $0.3-2.0$\,keV bandpass for \textit{Chandra} and $0.3-1.2$\,keV
for \textit{XMM-Newton}.

If circular annuli are used, a blind radial profile including the
bulge emission in the \textit{Chandra} image yields a surface
brightness scale height of $\sim$2\,arcmin ($\sim$4\,kpc).  
The bulge and halo are separate components, with the bulge having a
smaller scale height and larger surface brightness.  This suggests a
larger scale height for the halo.  

As the halo may not be spherically symmetric, we can measure the
profile in $\hat{z}$ using regions parallel to the midplane of the
galaxy \citep[similar to the approach used for the NGC\,891 \textit{Chandra} data
in][]{strickland04}.  This profile is shown in
Figure~\ref{figure.midplane_profile}.  The image on top is the
adaptively smoothed \textit{Chandra} image, with the 6\,arcmin by
0.25\,arcmin boxes we used overlaid on the image.  The mean surface
brightness in each region is shown in the bottom panel, and is blindly
corrected for the point source masks and other unexposed areas by
scaling to the nominal area in each box.

The \textit{Chandra} data are shown as the black boxes in the bottom
panel of Figure~\ref{figure.midplane_profile}, and includes the bulge
emission.  To derive the scale height, we fit an exponential profile
($I(z) = I_0e^{-z/z_0}+C$), finding $z_0 \sim 0.6$\,arcmin (1.3\,kpc) for the
\textit{Chandra} data.  This is similar to that in
\citet{strickland04} based on their
encircled light values, and somewhat smaller than the \textit{ROSAT}
value reported by \citet{bregman94}.  The \textit{XMM-Newton} values
are shown as red triangles in Figure~\ref{figure.midplane_profile} and
result in a slightly larger $z_0 \sim 0.9$\,arcmin (2\,kpc).  As these
values are extracted in wider boxes (6\,arcmin by 0.5\,arcmin) because 
of the large point source masks, we expect a bias towards higher $z_0$.
On the other hand, the \textit{XMM-Newton} images mask most of the bulge
emission (under the ULX point source mask) that is included in the
\textit{Chandra} fit.  Since the spectrum of the bulge emission shows
that it is clearly a different physical component (which we expect to have
a smaller scale height), the \textit{Chandra} value quoted above is
probably too small.  Since the bulge emission is not entirely masked
in the \textit{XMM-Newton} image, we can improve this value by ignoring
a region within about an arcminute of the galactic center.  We do this by
extracting profiles in 3\,arcmin by 0.5\,arcmin north and south of the
bulge in the \textit{XMM-Newton} data, seen as green diamonds and blue 
stars in Figure~\ref{figure.midplane_profile} respectively. 

The similarity between the values for the long boxes and the northern
shorter boxes in Figure~\ref{figure.midplane_profile} shows that the
correction is small, with the dominant component being the bright emission
near the disk on the north side.  The halo emission itself is asymmetric
both north-to-south and east-to-west.  The north--south asymmetry is
well known and can be easily seen in Figure~\ref{figure.midplane_profile}
where the boxes are centered on the galactic center.  The
east--west asymmetry can be seen in the horizontal dashed and dotted
lines in Figure~\ref{figure.midplane_profile}, which represent a
constant ``background'' in the fit to the \textit{Chandra} data.  
Despite nearly identical scale heights on either side of the galaxy,
the west side is decidedly brighter out to the extent of the S3 chip.  
Emission in the wings of the PSF of bright subtracted point sources can 
account for less than 20\% of the discrepancy.  A cluster of faint
point sources cannot be ruled out. 

On the north side of the bulge, we find $z_0 \sim 1$\,arcmin (2\,kpc) east of
the galaxy and $z_0 \sim 0.6$\,arcmin (1.3\,kpc) on the west side.  In contrast,Z
the surface brightnesses obtained on the south side of the bulge
indicate scale heights of $z_0 \sim 1.5$\,arcmin (3.2\,kpc) east of the disk and
$z_0 \sim 1.8$\,arcmin (3.9\,kpc) to the west.  The difference in these values
suggests that the inner halo may be a separate physical component.

However, the scale height is very sensitive to the innermost surface
brightness values.  These values are the most affected by the point
source masks (which are more concentrated near the disk), the bulge,
and the residual 10\% of the ULX emission.  The change in surface
brightness across each extraction zone is also highest near the disk,
so our blind correction to the nominal area increases the uncertainty
in these bins far beyond the $\sqrt{N}$ errors plotted in
Figure~\ref{figure.midplane_profile}.  

Still, it is likely that the true scale height, at least in the outer
halo, is even larger than the $z_0 \sim 1.5$\,arcmin found in the
southern boxes in the \textit{XMM-Newton} data.  When we correct for
the bulge and ULX by subtracting model profiles (assuming a circular
PSF for the ULX and a circular model fit to the bulge in the
\textit{Chandra} data), $z_0$ increases to near 1.0\,arcmin in the
northern boxes (where the inner halo is bright) and 2.5\,arcmin in the
southern boxes.  This is because $z_0$ is highly sensitive to the mean
surface brightness in the innermost bins, which is overestimated
according to these corrections.  Overall, a scale height of $z_0
\gtrsim 2$\,arcmin ($\sim4$\,kpc) is likely in the outer halo.  The density
scale height is twice this value, or $H \gtrsim 8$\,kpc, which is
consistent with \citet{bregman94} considering they included all diffuse
emission.  In any case, a scale height below 1\,arcmin is strongly disfavored.

Finally, we note that there is a patch of diffuse emission to the
northwest of the galaxy with approximately a 12\,arcmin separation.
This is visible in both the \textit{Chandra} and \textit{XMM-Newton} data
and the emission is centered near 
$\alpha = $02h21m32.2s, $\delta = +$42d25m12s.  The emission is roughly
circular with a radius of about 2\,arcmin, but it has some
substructure with a brighter ``bar'' running roughly parallel to
NGC\,891.  The emission is near the \textit{Chandra} point source
CXOSEXSI\,J022142.6+422654 (which may also be 1AXG\,J022139+4226, a
source identified as 0.5\,arcmin away but in an error circle of
1.1\,arcmin).  We have not found any records in the literature of this
source despite its appearance in the ACIS-I chips of the archival
\textit{Chandra} data, and unfortunately the optical and IR data at
this position is limited.  In 2MASS and Palomar/DSS images, we see no
obvious signs of clustering.  An \textit{Xspec} {\sc apec} fit with
only Galactic absorption finds a good fit for $kT \sim 3$\,keV, so we
speculate that this is a galaxy cluster at unknown redshift.  The
$0.3-10$\,keV flux is $F_X \sim 1.3\times
10^{-13}$\,erg\,s\,cm$^{-2}$.  

\begin{figure}
\begin{center}
\includegraphics[width=0.75\linewidth,angle=90]{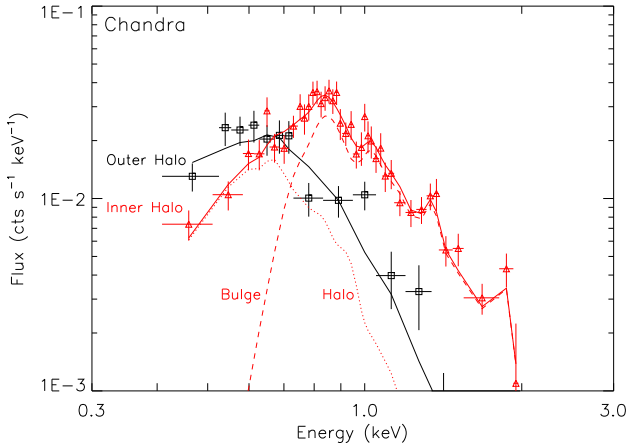}
\caption{\textit{Chandra} inner (red) and outer (black) halo spectra with 
model components for the inner halo spectrum overplotted.  The data have been
rebinned to 5$\sigma$ $S/N$ per bin for plotting purposes.  The outer halo
is fit with a single absorbed {\sc apec} model.}
\label{figure.chandra_spectra}
\end{center}
\end{figure}

\subsection{Source Spectra}

We extracted inner and outer halo spectra from the regions described
above in point-source subtracted, exposure-corrected data.  The
\textit{Chandra} spectra were separately extracted from obsIDs 794 and
4613 and combined, with corresponding response files generated.  We
likewise extracted background spectra from regions near the edge of
the ACIS-S3 chip.  The \textit{XMM-Newton} spectra were extracted from
analogous regions, and we generated their response files using the
XMM-ESAS scripts {\it mos-spectra}, {\it pn-spectra}, {\it mos\_back},
and {\it pn\_back}.  All of the spectra were binned to 25\,counts per
bin for fitting.  We restrict the fitting bandpass in the
\textit{Chandra} data from $0.4-2.0$\,keV.  Above 2.0\,keV, there is
little source flux and the data do not constrain any fit.  Likewise,
the \textit{XMM-Newton} data are fitted between $0.3-3.0$\,keV in the
MOS and $0.4-3.0$\,keV in the pn.  The lower bounds are determined by
the rise of the low-energy particle background in each detector.

We use {\it Xspec} v12.7.1 for spectral fitting \citep{arnaud96}.  We
use $\chi^2$ as the goodness-of-fit parameter, and define an
acceptable model as one that is not ruled out at 95\% confidence based
on the $\chi^2$ value.  Errors on the model parameters are determined
using the {\it Xspec} \textit{steppar} task, and model fluxes are
quoted for a $0.3-3$\,keV bandpass.

The diffuse source emission is made up of two or more components: the
bulge and (inner/outer) halo.  We begin with the simplest case of a
homogenous halo, but the inner and outer halo may be produced by a
galactic fountain and IGM accretion respectively, and so may have
different properties.  The halo has an X-ray spectrum that peaks near
0.6\,keV, whereas the hotter bulge has a spectrum peaking near
0.9\,keV (Figure~\ref{figure.chandra_spectra}).  The bulge emission is
unimportant in the outer halo, but contributes near the disk.  As we
expect these to be thermal plasmas in collisional ionization
equilibrium, our default model is {\sc apec/vapec} \citep{smith01}.
Since photoelectric absorption is important, as an absorption model we
adopt {\sc TBabs} with the abundances of \citet{wilms00}. 
All fits incorporate the Galactic absorption column of
$N_H = 6.5\times 10^{20}$\,cm$^{-2}$ \citep{kalberla05} in addition to
variable absorption near the disk.  The influence of these choices (by
comparison with alternatives) is described in
Section~\ref{section.metallicity}.

The ULX in the \textit{XMM-Newton} data contributes approximately
55,000 counts in the pn and 17,000 in each MOS camera in the
$0.3-10$\,keV bandpass, peaking near 1\,keV.  Although we
mask the ULX to the 90\% encircled energy radius, even outside this it
contributes thousands of counts.  We adopt as a ULX model the best
fitting simple model from \citet{hodges-kluck12}: absorbed emission
from a hot accretion disk.  This model, {\sc TBabs*diskpbb}, is an
excellent fit to the ULX spectrum ($\chi^2 = 1169.1$/1170 d.o.f. for
the combined MOS and pn spectra).  We freeze the model parameters to
their best-fit values: $N_H = 2.3\times 10^{21}$\,cm$^{-2}$,
$T_{\text{in}} = 1.62$\,keV, and $p = 0.54$ \citep[$p$ is a free
parameter and $T(r) \propto r^{-p}$;][]{mineshige94}.  This model is
included in all our \textit{XMM-Newton} fits with the normalization
allowed to vary, but we find that it reproduces the expected
contribution of the ULX flux very well in annuli around the ULX.  The
ULX spectrum does vary slightly with radius because the PSF is energy
dependent.  However, for the relevant energies, the expected
difference is approximately 1\%, which is smaller than the Poisson
uncertainty in the ULX flux in the source extraction regions.

\subsection{X-ray Background}

\begin{figure*}
\begin{center}
\includegraphics[width=0.45\linewidth]{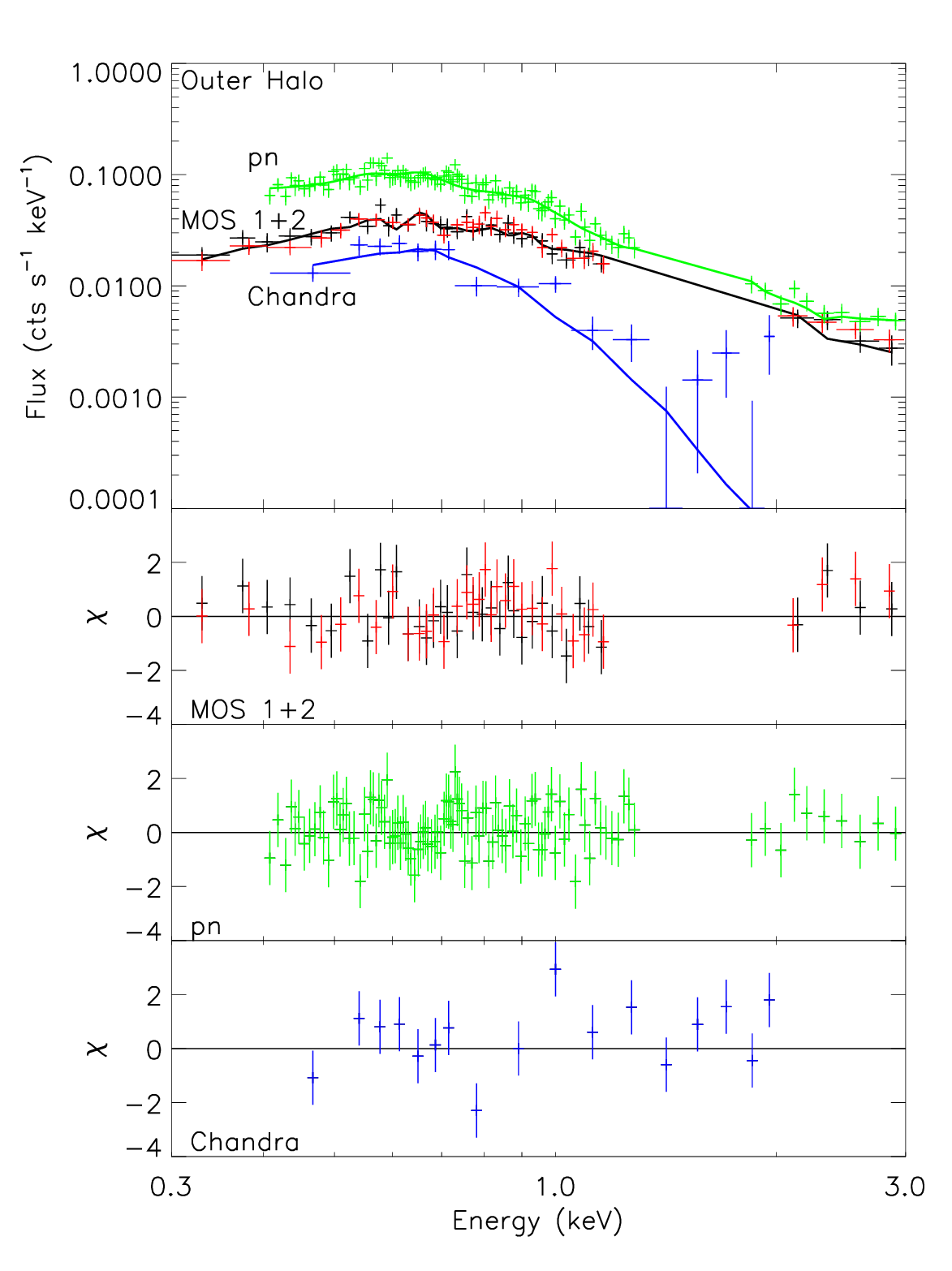}
\includegraphics[width=0.45\linewidth]{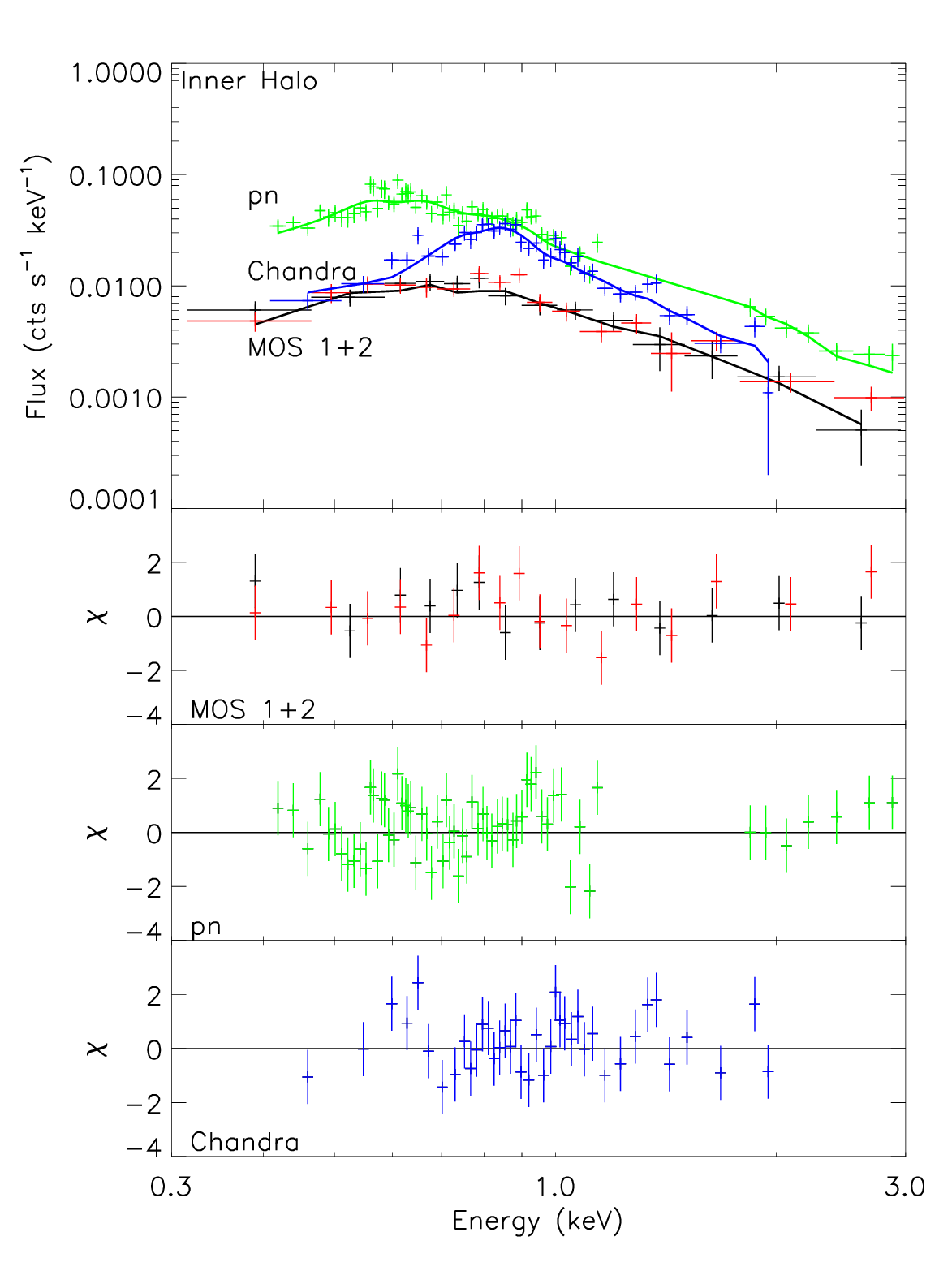}
\caption{\textit{XMM-Newton} and \textit{Chandra} spectra of the outer
  halo (Left) and inner halo (Right), with the MOS 1$+$2 in black and
  red, pn in green, and \textit{Chandra} ACIS in blue.  The best fits
  are overlaid with their residuals below, and the data have been
  rebinned to 5$\sigma$ for plotting purposes.  Note that the
  \textit{Chandra} inner halo spectrum contains much more bulge
  emission than the \textit{XMM-Newton} spectra where it is mostly
  excised with the brightest ULX emission.}
\label{figure.halo_spectra}
\end{center}
\end{figure*}

The X-ray background includes instrumental noise, particle
backgrounds, the cosmic X-ray background, the contribution from the
Local Bubble, and the hot Galactic halo.  The latter two are
particularly important for our purposes because the Galactic halo has
a similar temperature ($kT\sim0.25$\,keV) to that of NGC\,891, and the
Local Bubble contributes to the emission lines by which we determine
the temperature of the source plasma.  Such emission is typically
subtracted by using a scaled spectrum extracted from nearby
source-free regions.  However, the emission from NGC\,891's halo
covers most of the ACIS-S3 chip and the central chips of the MOS
detectors, and the true scale height of the halo is unknown.
The \textit{XMM-Newton} particle background also induces strong
fluorescent lines (the Al and Si K$\alpha$ lines at 1.49\,keV and
1.75\,keV are the important lines in our fitting bandpass) that are
highly variable with position such that scaled background subtraction
leaves large P\,Cygni type profiles in the $1-2$\,keV range.  Thus,
the background regions for the \textit{Chandra} data contain an
unknown amount of source flux, while those for the MOS detectors
(placed farther out) do not directly subtract the instrumental
background for the chip of interest.  Our analysis fits use the
background-subtracted spectra, but we perform parallel fits using
background subtracted spectra and raw spectra where most of the
background is fit explicitly \citep[with the unexposed chip
backgrounds subtracted, following for the \textit{Chandra} and
\textit{XMM-Newton} data respectively][]{markevitch03,snowden04}.


\begin{deluxetable*}{llllllll}
\tablenum{3}
\tabletypesize{\scriptsize}
\tablecaption{1-T Halo Model Fits}
\tablewidth{0pt}
\tablehead{
\colhead{Spectrum} & \colhead{$N_{H,1}$} & \colhead{$kT_1$} & \colhead{$Z_1$} &
\colhead{$N_{H,2}$} & \colhead{$kT_2$} & \colhead{$Z_2$} &
\colhead{$\chi^2$}\\ 
& \colhead{($10^{20}$\,cm$^{-2}$)} & \colhead{(keV)} & \colhead{($Z_{\odot}$)} &
\colhead{($10^{20}$\,cm$^{-2}$)} & \colhead{(keV)} & \colhead{($Z_{\odot}$)}
}
\startdata
\multicolumn{8}{c}{\sc Outer Halo ($z>2.8$\,kpc)}\\
\cline{1-8}\\
Chandra   & 6.5(f) & $0.24_{-0.03}^{+0.07}$ & $0.04_{-0.02}^{+0.05}$ & - & - & - & 102.6 (105)\\
MOS 1$+$2 & 6.5(f) & $0.21\pm0.02$          & $0.11_{-0.04}^{+0.05}$ & - & - & - & 164.5 (180)\\
pn        & 6.5(f) & $0.21_{-0.02}^{+0.03}$ & $0.08_{-0.03}^{+0.05}$ & - & - & - & 166.1 (175)\\
Joint     & 6.5(f) & $0.21\pm0.02$          & $0.08\pm0.03$          & - & - & - & 426.2 (460)\\
\cline{1-8}\\
\multicolumn{8}{c}{\sc Inner Halo ($0.6<z<2.8$\,kpc)}\\
\cline{1-8}\\
Chandra   & $10_{-4}^{+20}$ & $0.22\pm0.02$ & $0.12_{-0.06}^{+0.10}$ & $77_{-10}^{+12}$ & $0.58_{-0.03}^{+0.06}$ & 1.0(f) & 74.0 (79)\\
MOS 1$+$2 & $24_{-18}^{+27}$& $0.19\pm0.05$ & $0.10_{-0.07}^{+0.08}$ & 77(f) & 0.58 (f) & 1.0(f) & 155.5 (152)\\
pn        & $18_{-12}^{+22}$& $0.19\pm0.04$ & $0.11_{-0.07}^{+0.40}$ & 77(f) & 0.58 (f) & 1.0(f) & 117.2 (97)\\
Joint     & $16_{-7}^{+16}$ & $0.21\pm0.02$ & $0.11_{-0.06}^{+0.11}$ & $78_{-10}^{+11}$ & $0.58\pm0.05$ & 1.0(f) & 342.15 (328)
\enddata
\tablecomments{\label{table.halo_fits} The MOS and pn fits include a
  {\sc TBabs*diskpbb} component with all parameters except flux frozen
  as described in the text.  The fits corresponding to these models
  are shown in Figure~\ref{figure.halo_spectra}.  Errors are quoted at the 90\% confidence
  interval based on the \textit{Xspec} task \textit{steppar}, and (f)
  designates a frozen parameter.}
\end{deluxetable*}

In these fits, we explicitly model the astrophysical backgrounds
following the XMM-ESAS recipes.  We use the HEASARC X-ray Background
Tool\footnote{http://heasarc.gsfc.nasa.gov/docs/heasarc/xrayback.html}
to estimate the background surface brightness from the \textit{ROSAT}
All-Sky Survey (RASS) data.  This tool also provides a spectrum and
response file, which we use to obtain the flux in each component of
the background.

With this tool, one usually defines a large region around the object
of interest, assuming that the background is approximately correct for
the line of sight.  However, there are a few bright diffuse sources
within 30\,arcmin of NGC\,891 (most notably the galaxy cluster
containing 3C\,66B to the north).  Thus, we measure the background in
a circle with radius 30\,arcmin from an off-position to the southwest.
We use the flux in the R45 and R67 bands ($0.44-2.04$\,keV).  We
convert the surface brightness to energy units via the HEASARC
WebPIMMS
tool\footnote{http://heasarc.gsfc.nasa.gov/TOols/w3pimms\_pro.html}
with a $0.1$\,keV {\sc apec} model for the Local Bubble, a $0.25$\,keV
absorbed {\sc apec} model for the hot Galactic halo, and an absorbed
power law ($\Gamma = 1.46$) for the cosmic X-ray background.  This
yields, in the $0.3-3.0$\,keV bandpass, a surface brightness of $I =
4\pm1\times 10^{-15}$\,erg\,s$^{-1}$\,cm$^{-2}$\,arcmin$^{-2}$.  As a
consistency check, we measured the surface brightness of the
background in a targeted 31.5\,ks \textit{ROSAT} PSPC observation of
NGC\,891 published by \citet{bregman94}.  We measured the $0.1-2.4$\,keV count rate
in a total of $250\pi$\,arcmin$^{-2}$ in the exposure-corrected field,
avoiding point sources and bright emission to the north and west.
This yields a surface brightness $I = 3\pm1\times
10^{-15}$\,erg\,s$^{-1}$\,cm$^{-2}$\,arcmin$^{-2}$ in the $0.3-3$\,keV band,
which is in agreement with the RASS value.

With the normalization in hand, we fitted the RASS spectrum with three
components: the Local Bubble (a thermal component with $kT =
0.1$\,keV), the Galactic halo (an absorbed {\sc apec} component with
$kT = 0.25$\,keV), and the cosmic X-ray background (an absorbed power
law with $\Gamma = 1.46$).  Although these values may
not be exactly correct, there are only a few spectral bins, so we freeze
them at the values quoted in the XMM-ESAS manual.  We use the Galactic absorbing column of
$N_H \approx 6.5\times 10^{20}$\,cm$^{-2}$ for the Galactic halo component.  In the $0.3-3$\,keV
bandpass, the fit finds that the Local Bubble contributes $\sim$6\% of
the flux, the (absorbed) Galactic halo $\sim$34\%, and the cosmic
X-ray background $\sim$56\%, which we use to determine the
normalizations of the model components in the source regions.

To verify that the \textit{Chandra} and \textit{XMM-Newton} background
spectra do not have unknown components, we fitted the same model
(adding Gaussian components for Al and Si K$\alpha$) but allowed the
parameters to float.  We find generally good agreement, but the
fractional contribution of the Local Bubble is poorly constrained
because the spectrum of the $kT = 0.1$\,keV plasma drops rapidly above
0.3\,keV.  However, the $0.3-3$\,keV surface brightnesses are in
agreement with the \textit{ROSAT} values: $S \approx
4.4\times10^{-15}$\,erg\,s$^{-1}$\,cm$^{-2}$\,arcmin$^{-2}$ in the MOS, $S
\approx 3.7\times 10^{-15}$\,erg\,s$^{-1}$\,cm$^{-2}$\,arcmin$^{-2}$ in the
pn, and $S \approx
3.9\times10^{-15}$\,erg\,s$^{-1}$\,cm$^{-2}$\,arcmin$^{-2}$ in
\textit{Chandra}, where these model fluxes are quoted for $0.3-3$\,keV
but fitted in the bandpasses noted above.  These results suggest that
the NGC\,891 halo surface brightness does indeed drop below the
background on the scale of the ACIS-S3 chip and that the regions used
for background subtraction are appropriate.  Any overestimate of the
background flux must be slight, and there do not appear to be large
discrepancies between the detectors.

Since the RASS background surface brightness and spectrum appear
consistent with the other data, for our explicit background fits we
fix the total flux, scaled to the source region, in the X-ray
background.  In Section~\ref{section.metallicity}, we show that fits
with background subtraction and fits explicitly fitting the background
are in good agreement. 

\begin{figure*}
\begin{center}
\includegraphics[width=0.3\linewidth,angle=90]{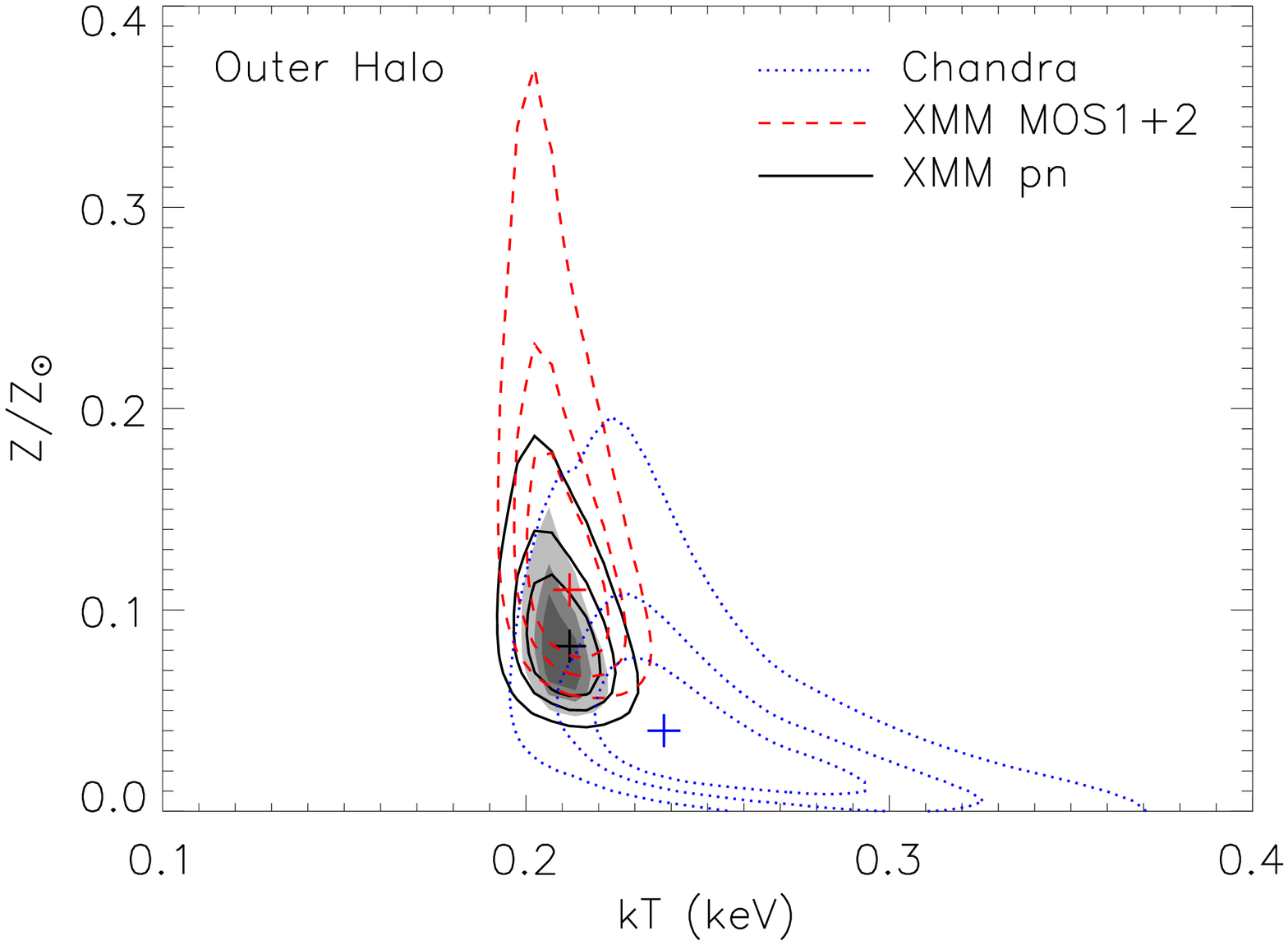}
\includegraphics[width=0.3\linewidth,angle=90]{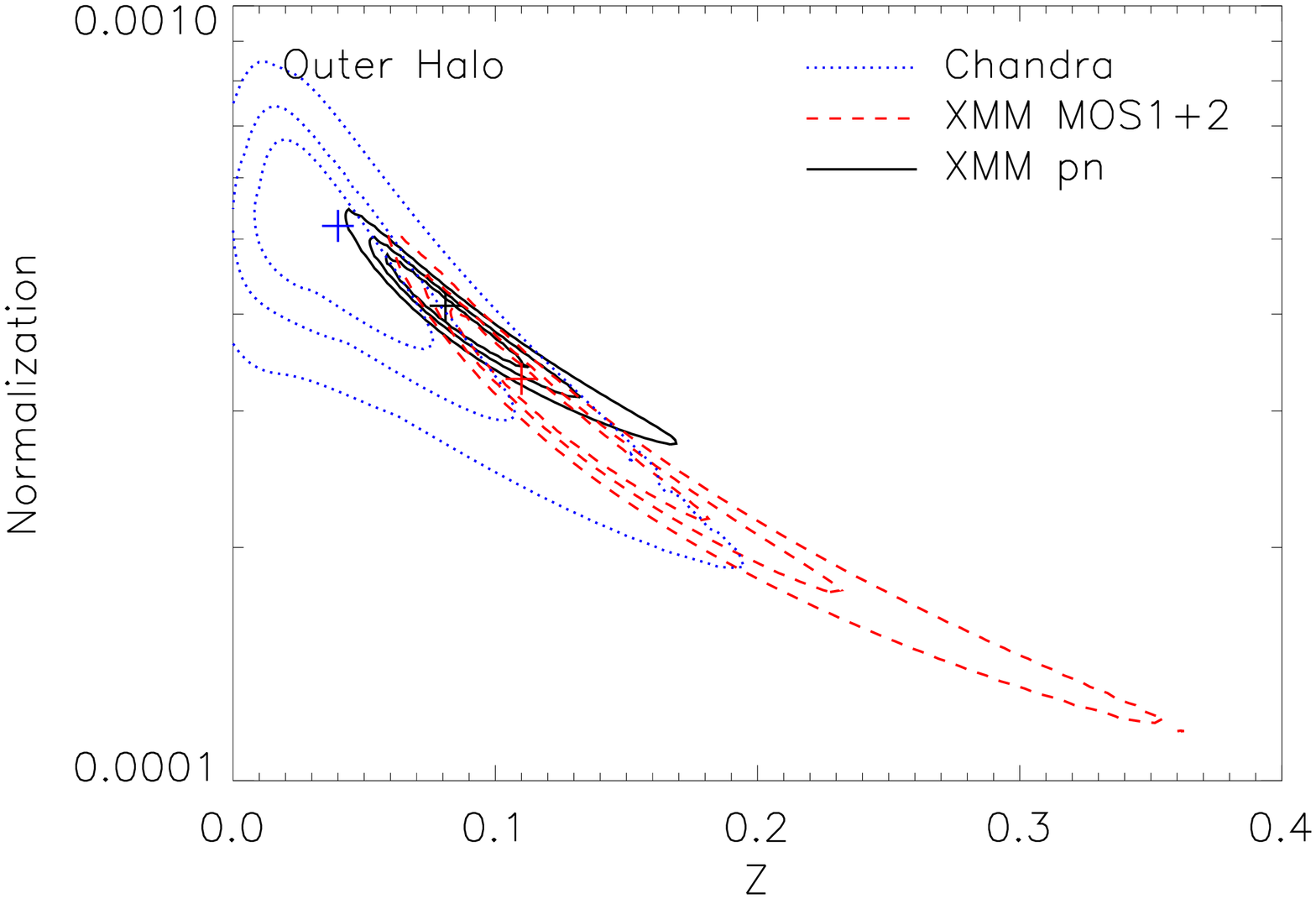}
\includegraphics[width=0.3\linewidth,angle=90]{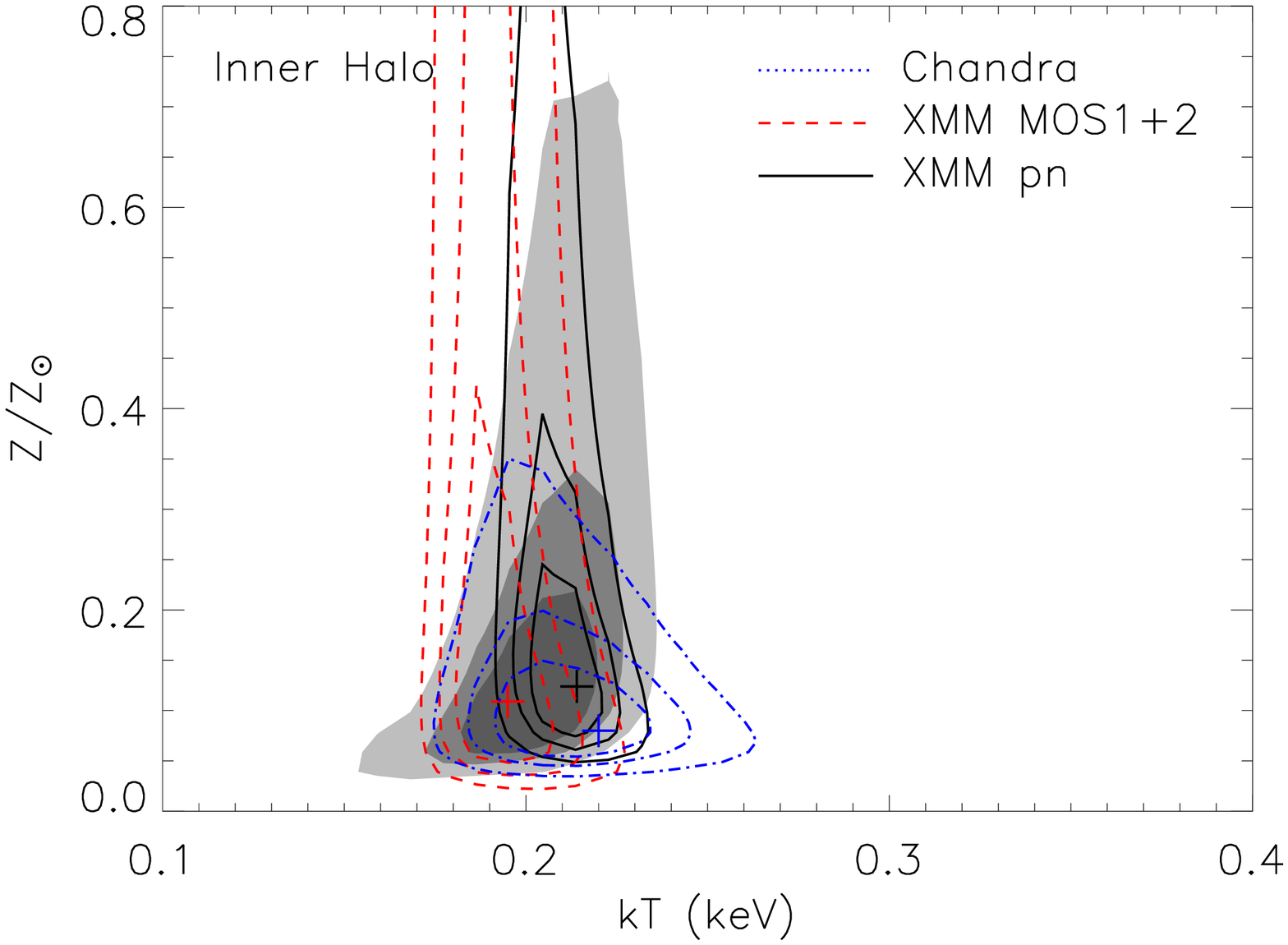}
\includegraphics[width=0.3\linewidth,angle=90]{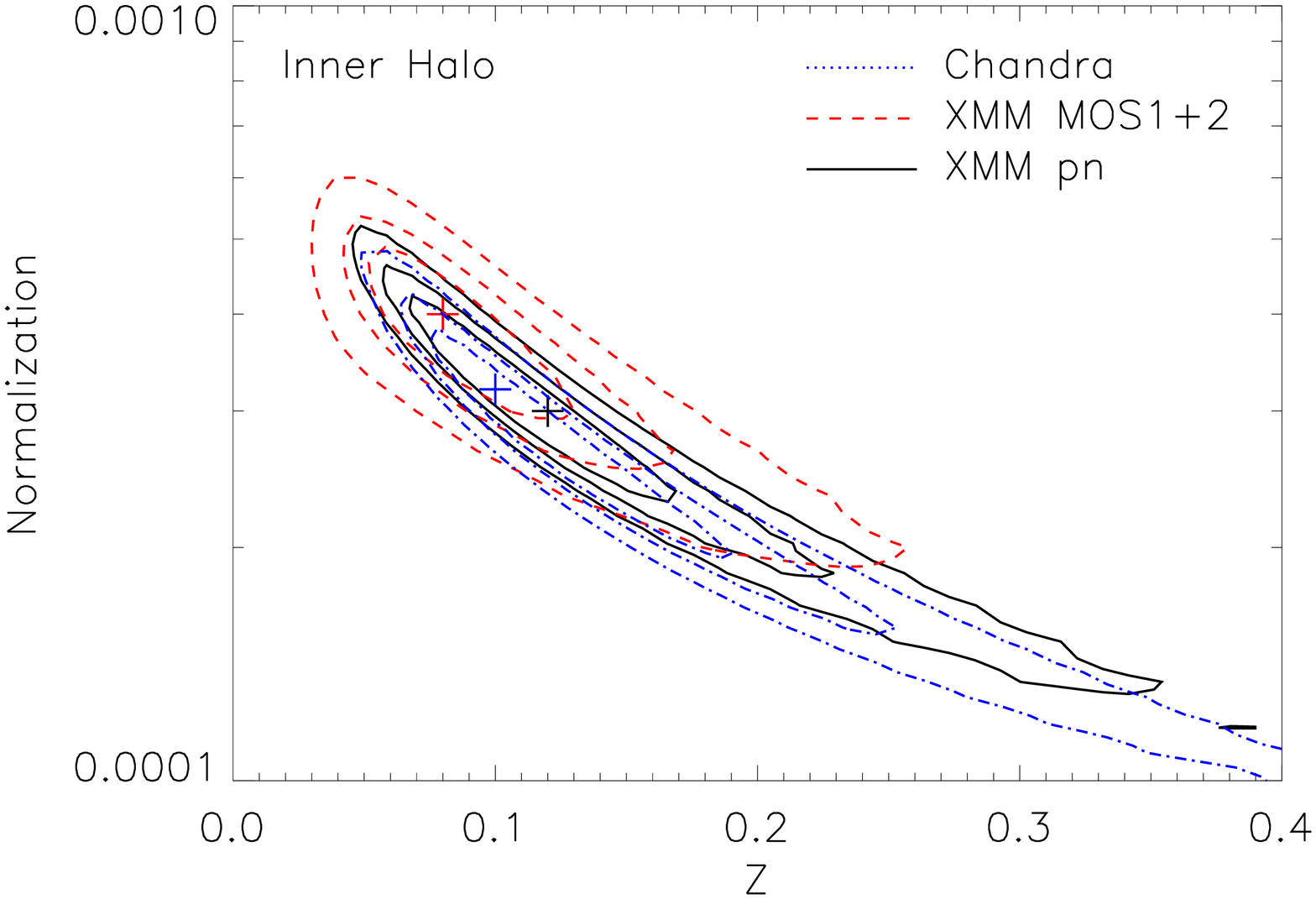}
\caption{Left: $Z$ vs. $kT$ contours for the outer halo (top) and
  inner halo (bottom).  The contours represent 1, 2, and 3$\sigma$
  deviations from the best-fit values denoted by crosses.  The shaded
  regions show the contours for the joint fit.  Right: As left, for
  the flux (normalization) vs. $Z$ contours.  The normalizations have
  been scaled to the \textit{Chandra} value (due to different areas)
  and the absolute value is unimportant.}
\label{figure.halo_contours}
\end{center}
\end{figure*}

\section{Spectral Fits \& Halo Metallicity}
\label{section.metallicity}

In this section we present our model fits.  Unless stated otherwise,
we refer to the background-subtracted spectra.  First, we describe
single-temperature (1-T) fits to the outer and inner halo, followed by
a more complex two-temperature (2-T) model.  We then compare the
results of our 1-T fits to the fits obtained by fitting the background
explicitly.  Finally, we discuss several potential sources of
systematic error and the effect on our results.

\subsection{Models}

The models we use for the different spectral components are described
in the previous section and summarized here.

The outer halo
is represented by an absorbed thermal ({\sc TBabs*apec}) model, with
$N_H$ fixed at the Galactic value.  This model represents emission
from a hot plasma in collisional ionization equilibrium modified by
photoelectric absorption by an intervening column of atomic gas.  The
inner halo is the same, but with $N_H$ variable.  The bulge uses the
same model, but with $Z = Z_{\odot}$.  For the ULX, we use an absorbed
accretion disk model ({\sc TBabs*diskpbb}) as in
\citet{hodges-kluck12} and described above.  Following
\citet{snowden04}, we model the Local Bubble with an {\sc apec} model
of $kT = 0.1$\,keV.  The Galactic halo and cosmic X-ray background are
modeled as {\sc TBabs(apec+pow)}, where $N_H$ is set to the Galactic
value, $kT = 0.25$\,keV, and $\Gamma = 1.46$.  The total flux is
scaled to the RASS value, and the power in each component is
constrained by simulaneously fitting the RASS spectrum.  The Al and Si
K$\alpha$ lines are fit by {\sc gauss} components, with fixed energy
and $\sigma = 0$ (instrumental linewidth).

Thus, the 1-T model to the \textit{Chandra} outer halo is just {\sc
TBabs*apec}, whereas in \textit{XMM-Newton} data it is {\sc
TBabs*apec+TBabs*diskpbb}.  In the inner halo, we add the bulge
component.  In the fits where the X-ray background is explicitly fitted,
we add {\sc apec+TBabs(apec+pow)} components and Gaussians in the 
\textit{XMM-Newton} data.  The best fit values for the
background-subtracted fits are tabulated in 
Table~\ref{table.halo_fits}.

\subsection{Outer Halo}

The \textit{Chandra} spectrum (Figure~\ref{figure.halo_spectra}) is
fit well by a thermal model with $kT = 0.24_{-0.03}^{+0.07}$\,keV
and $Z = 0.04_{-0.02}^{+0.05} Z_{\odot}$.  The MOS and pn spectra
prefer a slightly cooler ($kT \approx 0.21$\,keV;
Table~\ref{table.halo_fits}) plasma with a metallicity $Z \sim 0.1
Z_{\odot}$, but the results agree within their 90\% error bars (for a
single parameter of interest).  This is cooler than the 0.3\,keV found
in the \textit{ROSAT} data by \citet{bregman94}, although most of the
light in their spectrum comes from the inner halo region.  Likewise,
it is slightly cooler than the 0.26\,keV \textit{XMM-Newton} value
reported by \citet{temple05} using a much shorter exposure, but again
their spectrum is dominated by the inner halo.  All of the fits are
excellent (Figure~\ref{figure.halo_spectra}) and the temperatures are
constrained to 15\%.


\begin{deluxetable*}{llllll}
\tablenum{4}
\tabletypesize{\scriptsize}
\tablecaption{2-T Model Fits}
\tablewidth{0pt}
\tablehead{
\colhead{Spectrum} & \colhead{$kT_1$} & \colhead{$kT_2$} &
\colhead{Absorbed $0.3-3.0$\,keV} & \colhead{Norm. Ratio} & \colhead{$\chi^2$}\\
& \colhead{(keV)} & \colhead{(keV)} & \colhead{$F_2/F_1$} & \colhead{EM$_2$/EM$_1$}
}
\startdata
\multicolumn{6}{c}{\sc Outer Halo ($z>2.8$\,kpc)}\\
\cline{1-6}\\
Chandra   & $0.09_{-0.02}^{+0.06}$ & $0.29\pm0.05$          & $1.9\pm0.7$ & $0.33\pm0.07$ & 109.4 (105) \\
MOS 1$+$2 & $0.12_{-0.03}^{+0.04}$ & $0.25_{-0.02}^{+0.05}$ & $3.4\pm0.3$ & $0.29\pm0.06$ & 173.2 (180) \\
pn        & $0.11\pm0.02$          & $0.27\pm0.03$          & $2.5\pm0.3$ & $0.37\pm0.07$ & 169.1 (175)\\ 
Joint     & $0.11\pm0.02$          & $0.27\pm0.03$          &           - & - & 436.5 (460)\\
\cline{1-6}\\
\multicolumn{6}{c}{\sc Inner Halo ($0.6<z<2.8$\,kpc)}\\
\cline{1-6}\\
Chandra   & $0.09_{-0.04}^{+0.14}$ & $0.25\pm0.05$          & $5\pm2$     & $0.3\pm0.1$  & 72.2 (79) \\
MOS 1$+$2 & $0.13\pm0.05$          & $0.26\pm0.06$          & $2.6\pm0.4$ & $0.5\pm0.1$ & 152.3 (152)\\
pn        & $0.05_{-0.03}^{+0.06}$ & $0.21\pm0.05$          & $3.1\pm0.7$ & $0.4\pm0.2$ & 111.9 (97) \\
Joint     & $0.09_{-0.03}^{+0.05}$ & $0.24\pm0.03$          & - & - & 346.4 (328) 
\enddata
\tablecomments{\label{table.fountain_fits} The fits include components
  as in Table~\ref{table.halo_fits}, including the ULX in the
  \textit{XMM-Newton} data and a thermal bulge component in the inner
  halo.  Errors are quoted at the 90\% confidence interval for one
  parameter of interest based on the \textit{Xspec} task
  \textit{steppar}, and for each $kT$ value the other is frozen during
  the error search to prevent reversals.  Column 4 is the $0.3-3$\,keV
absorbed flux ratio between the two thermal components, and column 5
is the ratio between their model normalizations.}
\end{deluxetable*}

The higher signal and higher spectral resolution of the
\textit{XMM-Newton} data also mitigates the degeneracy between $kT$
and $Z$ in the \textit{Chandra} data, as seen in a 2-parameter $\Delta
\chi^2$ contour plot (Figure~\ref{figure.halo_contours}).  It is
evident from this plot that solar metallicity is ruled out at more
than 3$\sigma$ confidence in each detector.  There is an important
degeneracy between the flux and metallicity (the normalization values
have been scaled to the \textit{Chandra} value due to different
extraction areas, and the absolute values are unimportant), but
subsolar metallicity is preferred as strongly as in the $Z-kT$ plot,
where the degeneracy between the parameters (which manifests as
non-zero slope) is much smaller.  Since the results from each detector
are largely in agreement (Table~\ref{table.halo_fits}), we can jointly
fit all four spectra with a 1-T model.  With a
total of 24602 source counts in the fitting bandpass (of which we
estimate fewer than $\sim$6000 are ULX photons), we find 5$\sigma$
limits of $kT = 0.20_{-0.02}^{+0.05}$\,keV and $Z =
0.11_{-0.06}^{+0.31} Z_{\odot}$ for a single parameter of interest.
 
We measure the relative abundances of O and Fe by using 
a thermal model where the abundances of each metal are variable ({\sc
  vapec}).  Since all of the elements
contribute to the continuum while only some have strong lines in the
soft X-ray bandpass, we allow only O and Fe to vary, tying the
abundances of like elements to their values and setting the rest to
the best-fit $Z$ value in Table~\ref{table.halo_fits} for each
spectrum.  We note that if the other elemental abundances are tied
together but allowed to vary, the best-fit O/Fe value is very similar
to that obtained when these abundances are frozen. 

\begin{figure}
\begin{center}
\includegraphics[width=0.75\linewidth,angle=90]{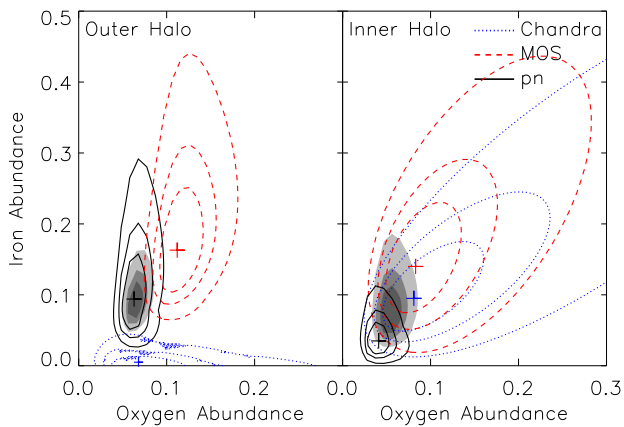}
\caption{2-parameter contour plots for O/Fe in the outer halo (Left)
  and inner halo (Right).  The contours are 1, 2, and 3$\sigma$
  contours and the minima are marked by crosses.  The shaded region is
  the 1, 2, and 3$\sigma$ contours for the joint fits.}
\label{figure.abund}
\end{center}
\end{figure}

The MOS and pn spectra indicate O/Fe$\sim$0.7, whereas the
\textit{Chandra} spectrum is consistent with $A_{\text{Fe}} = 0$, so
O/Fe is greater than unity but poorly constrained.  The fit contours
are shown in Figure~\ref{figure.abund}.  As in the {\sc apec} fits
(Figure~\ref{figure.halo_contours}), the MOS has a higher overall
metallicity than the pn, but the ratio of O/Fe is similar between
the two ($0.69$ in the MOS and $0.67$ in the pn, well within the
statistical error).  The \textit{Chandra} spectrum prefers a lower Fe
abundance because of the rather low flux between $0.7-1.0$\,keV
(Figure~\ref{figure.halo_spectra}), but the value is poorly
constrained, so the joint fit settles on a value near O/Fe$\sim$0.7
(Figure~\ref{figure.abund}).  However, O/Fe is quite sensitive to
temperature (higher temperatures tend to depress $A_{\text{Fe}}$
and increase $A_{\text{O}}$).  The
error bars should therefore be considered to be somewhat larger than
the contours in Figure~\ref{figure.abund}.

\subsection{Inner Halo}

Despite its higher surface brightness, the inner halo contributes far
fewer source counts due to the smaller regions we have demarcated.
The \textit{XMM-Newton} spectrum is also more contaminated by the ULX
and both the \textit{Chandra} and \textit{XMM-Newton} spectra include 
contamination from the bulge emission (seen especially in the 
\textit{Chandra} spectrum in Figure~\ref{figure.chandra_spectra}).  
Contamination from the disk is minimal because of our extraction region
boundaries (outside two disk scale heights) and the very high absorption
towards the disk \citep[$N_H \sim 10^{22}$\,cm$^{-2}$ in][]{oosterloo07}.  
Finally, $N_H$ varies across the region, so it must remain a free parameter
in spectral fits.

We adopt a model analogous to the outer halo where we include a bulge
component and allow the $N_H$ column to vary above the Galactic
value.  As the bulge is nearly cospatial with the ULX in the
\textit{XMM-Newton} data, most of its emission is masked along with
the ULX.  The remaining bulge contribution is dominated by the Fe
L-shell lines between $0.7-1.0$\,keV, which can be fit by a range of
models with high absorption and moderate temperature.  We therefore
constrain the bulge temperature using the \textit{Chandra} data.  A
1-T model is a poor fit to the \textit{Chandra} data, so we use a 2-T
fit with the bulge metallicity frozen at solar (the metallicity is
poorly constrained, but consistent with solar and well above the outer
halo value).  This yields $kT \sim 0.6$\,keV
(Table~\ref{table.halo_fits}).  For the bulge contribution in the MOS
and pn, we use a thermal component with the absorption and temperature
frozen at the \textit{Chandra} values and the flux free to vary.
However, fits with unrealistic bulge fluxes (i.e., comparable to or
more than the total \textit{Chandra} bulge flux) are rejected.

\begin{figure}
\begin{center}
\includegraphics[width=1.0\linewidth]{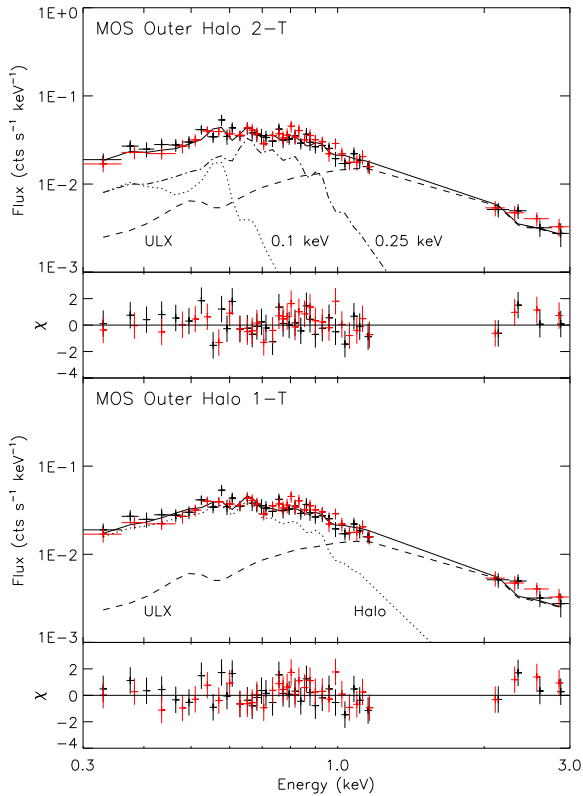}
\caption{Top: Folded EPIC MOS 1$+$2 outer halo spectrum fit by a 2-T
  galactic fountain model with components as tabulated in
  Table~\ref{table.fountain_fits}.  The dashed line shows the
  contribution of the ULX.  Bottom: As above, for the 1-T model shown
  in Table~\ref{table.halo_fits}.}
\label{figure.fountain_fit}
\end{center}
\end{figure}

\begin{figure}
\begin{center}
\includegraphics[width=1.0\linewidth]{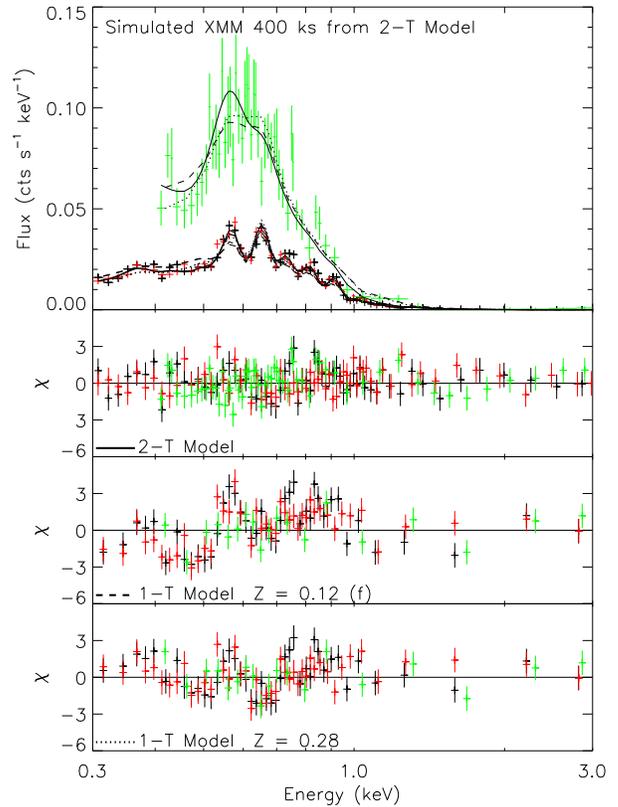}
\caption{Simulated 400\,ks \textit{XMM-Newton} spectra created from
  the best-fitting 2-T model (excluding the ULX).  The simulated
  spectrum is fitted with the 2-T (solid line) and 1-T (dashed line
  and dotted line) models, with the residuals for each below.  The
  middle residual panel shows a 1-T fit with $Z$ frozen at
  $0.12Z_{\odot}$ (as in Table~\ref{table.halo_fits}), and the bottom
  panel has variable $Z$ with a best-fit value of $Z =
  0.28Z_{\odot}$.  Note that the $y$-axis is linear rather than
  logarithmic, and that the data have been binned to 10$\sigma$
  significance for plotting purposes.}
\label{figure.fountain_sim}
\end{center}
\end{figure}

Using this model, we find good fits for thermal models in
the EPIC spectra that are consistent with the \textit{Chandra} data
(Table~\ref{table.halo_fits}).  These fits and their residuals are
shown in Figure~\ref{figure.halo_spectra}.  The best-fit temperatures
are $kT \approx 0.21$\,keV, while the metallicity is near $Z \sim 0.1
Z_{\odot}$ in each spectrum.  Because of the lower signal, the
confidence intervals for these parameters are substantially wider than
in the outer halo (Figure~\ref{figure.halo_contours}).  We note that
it is possible to find good fits with different combinations of $kT$,
$Z$, and $N_H$ with only small differences in $\chi^2$, so while the
2-parameter contours are accurate for the best fit, the 3-parameter
region of parameter space with good fits is substantially wider.  As
might be expected, the temperature is known much better than the
metallicity.  Still, it is noteworthy that the best-fit parameters in
each detector agree not only with each other (within the 90\% error
bars) but also with the outer halo values
(Table~\ref{table.halo_fits}).

As in the outer halo, we can improve the signal by jointly fitting the
four spectra.  This yields 3$\sigma$ limits of $kT =
0.21_{-0.06}^{+0.05}$\,keV and $Z = 0.11_{-0.07}^{+0.40} Z_{\odot}$.
The absorbing column is poorly constrained even in the joint fit, with
a 3$\sigma$ limit of $N_H = 16_{-7}^{+26} \times 10^{20}$\,cm$^{-2}$
on top of the Galactic value.  Based on the \ion{H}{1} data
(cf. Figure~\ref{figure.HI_halo}), we expect a value above $N_H =
10^{21}$\,cm$^{-2}$ but below $10^{22}$\,cm$^{-2}$, although generally
we expect more emission to come from less absorbed regions, biasing
the fitted $N_H$ value toward lower values.  Thus, we cannot rule out
a higher metallicity for the inner halo than the outer halo, but both
are likely to be subsolar.

The signal is not sufficient to reliably measure the relative abundances if the bulge
component is also allowed to vary.  By freezing the bulge component at
the best-fit {\sc apec} values, we can obtain O/Fe values, but these
should be interpreted with caution.  We proceed as with the outer
halo, and Figure~\ref{figure.abund} shows the resultant contours.
Like in the outer halo, the best fits tend to have O/Fe$<1.0$,
although the pn now finds a value of O/Fe slightly smaller than 1.0.
The \textit{Chandra} spectrum indicates a higher Fe abundance than in
the outer halo, in agreement with the {\sc apec} fit
(Table~\ref{table.halo_fits}).  However, the uncertainty is large and
the bulge emission is quite important to this spectrum.  If we fit the
total \textit{Chandra} spectrum extracted from both the inner and
outer halo, the best-fit value is then O/Fe$=1.3_{-0.4}^{+0.3}$.
The 1$\sigma$ contour for
the joint fit is in agreement with the joint fit for the outer halo,
but the best fit finds a slightly higher O/Fe.  The picture is
confused by the bulge emission, but at face value the O/Fe value is
slightly higher in the inner halo than the outer halo.

The best-fit values suggest that the inner and outer halo are composed
of the same plasma despite the apparent connection between the inner
halo morphology and the activity in the disk.  ``Deprojecting'' the
\textit{Chandra} spectrum by using a scaled outer halo spectrum as the
background confirms that the inner halo temperature we derive is not
merely due to outer halo material in the line of sight.  Thus, the hot
halo of NGC\,891 is consistent with being an isothermal, $2-3\times
10^6$\,K plasma.  The agreement with the outer halo indicates that the
temperatures reported by \citet{bregman94} and \citet{temple05} of
0.3\,keV and 0.26\,keV respectively may be boosted by contamination
from the bulge.  However, they are consistent with the limits obtained
by fitting the spectra individually, and there is some degeneracy
between $N_H$, $kT$, and $Z$.

\subsection{A 2-T Model}

It is also possible to obtain a good fit with a 2-T model where the
metallicity of both components is fixed at solar (the bulge emission is treated as
above).  This fit is motivated by the galactic fountain scenario as
well as the many 2-T fits in the literature to the halos
\citep{strickland04} and disks \citep[e.g.][]{tyler04} of massive
late-type galaxies.  The fit requires the
first component to have $kT_1 \sim 0.1$\,keV and the second to have $kT_2
\sim 0.25$\,keV (parameters given in Table~\ref{table.fountain_fits}).
The ratio of the emission measure is weighted towards the cooler
component.

As in the 1-T model, the inner and outer halo spectra have parameters
consistent with each other.  The fits in
Table~\ref{table.fountain_fits} are carried out in the same way as in
the 1-T model except that we add a second {\sc apec} component and fix
the ULX flux in the \textit{XMM-Newton} spectra to the best-fit value
in the 1-T model.  The error on each temperature is computed while
freezing the other because they are separated by only $0.15$\,keV and
can exchange places in the parameter space search.

\begin{figure}
\begin{center}
\includegraphics[width=1.0\linewidth]{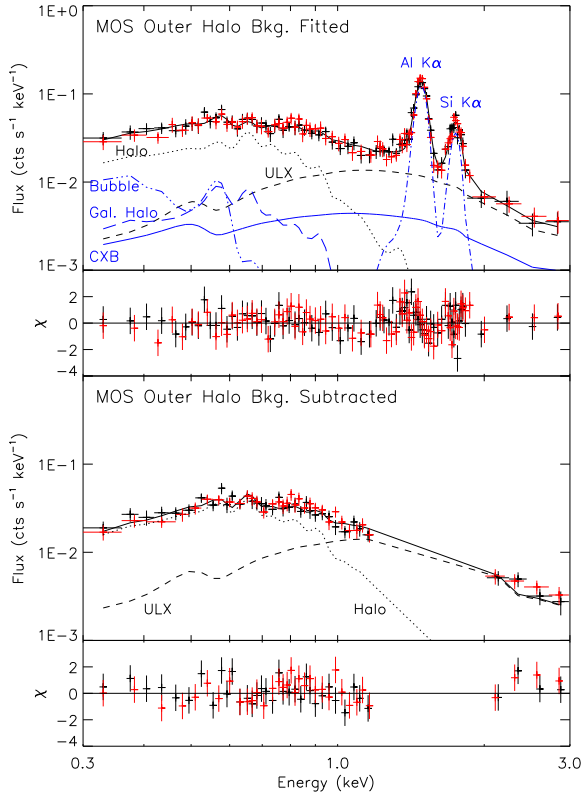}
\caption{Top: Folded EPIC MOS 1$+$2 outer halo spectrum with background components
explicitly fitted and residuals.  The dotted black line is the {\sc apec} model
for NGC\,891's halo, while the dashed black line is the ULX contribution.
The blue lines are background components: the Local Bubble (dash and dots), the
Galactic halo (dashed), the cosmic X-ray background (solid), and the instrumental
fluorescent lines (dash-dot).  The power in the X-ray background components is
constrained by the fit to the RASS spectrum.  
Bottom: As above, but for the background-subtracted (analysis) spectrum.}
\label{figure.bkgfit_bkgsub}
\end{center}
\end{figure}

\begin{figure}
\begin{center}
\includegraphics[width=0.75\linewidth,angle=90]{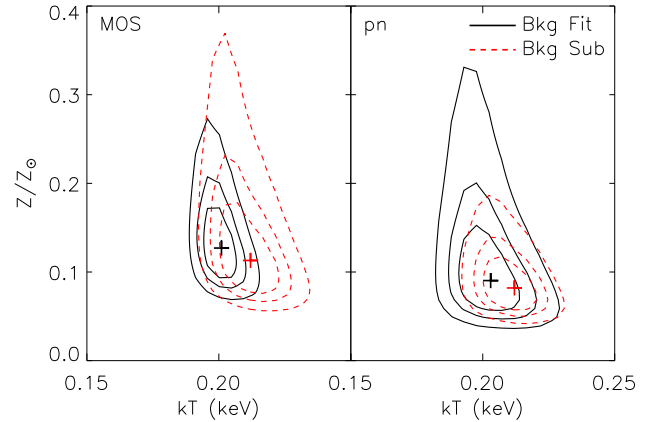}
\caption{Comparison of $Z - kT$ 1, 2, and 3$\sigma$ contours for the background-subtracted
  and background-fitted outer halo spectra for the MOS (Left) and pn
  (Right).  The solid black line represents contours from
  the background-fitted spectra and the dashed red line the
  background-subtracted (analysis) spectra.  Crosses identify the minima.}
\label{figure.bkgfit_bkgsub_contours}
\end{center}
\end{figure}

As the 1-T and 2-T models both have well constrained parameters and
produce good fits, it is worth asking how much complexity the spectrum
can distinguish and how much signal is required to rule out one of
these spectroscopically.  To examine this question, we used the
\textit{Xspec} tool \textit{fakeit} to simulate spectra based on the
best-fit 2-T model in the outer halo using the pn response, increasing
the signal by increasing the exposure time.  In these simulations, we
did not include the ULX.  We created 10 such spectra for each exposure
time, then fit them using the 1-T model.

We find that twice the $S/N$ ($\sim$400\,ks) is enough to distinguish
between the 1-T and 2-T models to high confidence.  The difference
between the models can be seen in the residuals
(Figure~\ref{figure.fountain_sim}), where the 1-T model underpredicts
the \ion{O}{7} flux and overpredicts the \ion{O}{8} flux.  Above
1.0\,keV, the models have similar residuals, and if the ULX were
included it would dominate the spectrum here.  Even at lower signal,
it may be possible to rule in favor of one model based on external
criteria or observable predictions of each model.  We examine a few of
these in Section~\ref{section.discussion}.  We also note, as in
\citet{strickland00b}, that the halo spectrum may not be simple and
could be undermodeled in either case.

The existing data are good enough to rule out non-thermal models
(i.e., those that do not produce emission lines).  The structure in
the spectrum below 1.0\,keV cannot be reproduced by realistic
absorption of a smooth continuum such as a power law.  Continuum
models produce poor fits with residuals similar to, but worse than,
those seen in Figure~\ref{figure.fountain_sim}.  

\subsection{Background Fits, X-ray Binaries, and Possible \ion{O}{7} Excess}

For the fits described above, we have used the background-subtracted
spectra.  To verify these fits, we also fitted the raw spectra (with
model quiescent particle backgrounds subtracted) while explicitly
fitting the X-ray background, following the XMM-ESAS guide \citep[and][for the \textit{Chandra} data]{markevitch03} 
and using the RASS data.  We also check the background fit to the source regions
against background spectra extracted for each detector.

The results in the outer halo for the MOS are shown in
Figure~\ref{figure.bkgfit_bkgsub}.  In the top panel, the background
components (folded through the MOS response) are overplotted in blue
while the source components are shown in black.  The spectrum is
fitted simultaneously with the RASS spectrum with the non-flux
parameters tied together to constrain the power in each component.
The overall flux in the background is computed from the surface
brightness measurements in the RASS and MOS background, so if the RASS
model (Local Bubble, Galactic halo, and cosmic X-ray background) is a
good description of the background, the source fits should be
identical to the background-subtracted case.
Figure~\ref{figure.bkgfit_bkgsub} demonstrates that both cases give
excellent fits, and Figure~\ref{figure.bkgfit_bkgsub_contours} shows
that both cases essentially give the same values for $Z$ and $kT$.
This is further verified by fitting the MOS and pn background spectra,
which are fit well by the RASS model.

\begin{figure}
\begin{center}
\includegraphics[width=0.75\linewidth,angle=90]{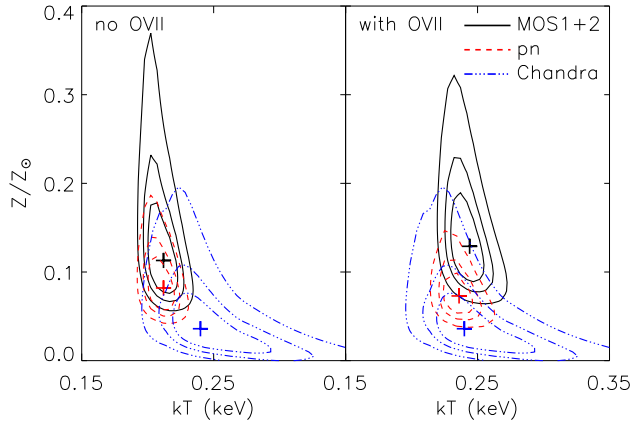}
\caption{Left: 1, 2, and 3$\sigma$ $Z-kT$ contours for fits to the \textit{XMM-Newton} and
\textit{Chandra} outer halo spectra as reported in
Table~\ref{table.halo_fits}.  Right: As left, but the
\textit{XMM-Newton} fits incorporate a Gaussian component at 0.58\,keV
representing the \ion{O}{7} complex.  The \textit{Chandra} contours
are the same as at left.  See text for discussion.}
\label{figure.o7_contours}
\end{center}
\end{figure}

\begin{figure}
\begin{center}
\includegraphics[width=1.0\linewidth]{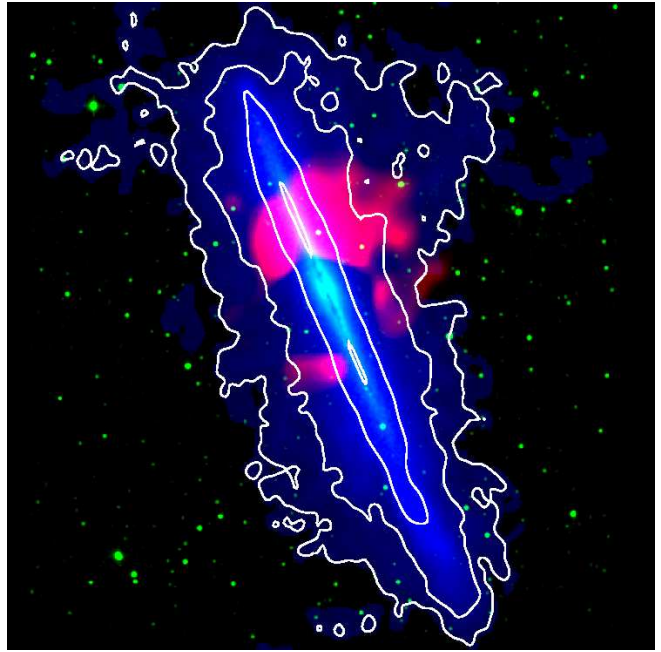}
\caption{Red: $0.5-0.6$\,keV adaptively smoothed \textit{XMM-Newton} image showing 
emission near the \ion{O}{7} complex.  Green: 2MASS $H$-band image.  Blue: 
Total \ion{H}{1} column density, with contours overlaid at $10^{19}$, $10^{20}$, $10^{21}$,
and $10^{22}$\,cm$^{-2}$.}
\label{figure.HI_OVII}
\end{center}
\end{figure}

The effect of the Local Bubble or Galactic halo on the measured
temperatures and metallicities must therefore be small, as must be the
contribution of the halo of NGC\,891 beyond the ACIS-S3 chip or
central MOS chips.  This is particularly important considering that
the Galactic halo has a temperature ($kT \sim 0.25$\,keV) similar to
NGC\,891's halo.

We have thus far not considered the flux contributed by X-ray binaries
(XRBs).  This flux may be from undetected sources or from the wings of
the PSF of detected ones.  Ignoring the XRB contribution can
artificially reduce the metallicity by forcing the thermal model to
boost continuum flux \citep[for a discussion,
see][]{kim03}.  The tell-tale sign of a missing XRB component in a
thermal spectrum is positive residuals at high energies.  The
residuals to our background-subtracted halo fits
(Figure~\ref{figure.halo_spectra}) do not indicate that an additional
component is required, except perhaps in the inner halo spectrum of
the pn.  The MOS and pn outer halo spectra show a similar structure in
the residuals above 2\,keV, but this appears to be a consequence of
the background subtraction (Figure~\ref{figure.bkgfit_bkgsub}).
As the \textit{XMM-Newton} spectra contain a ULX contribution, the
\textit{Chandra} data are a better indication of the influence of
XRBs.  Here, too, there is no obvious need for a harder component.
This is essentially what we would expect in the halo, since the XRB
population should predominantly occupy the disk. 

From a model perspective, no additional component is required (as with
the 2-T spectrum).  However, to see if an XRB component materially
improved the fit, we extracted an aggregate spectrum from the
brightest non-ULX disk sources and fit it with an absorbed power-law
model.  This yields a photon index of $\Gamma \sim 1.8$, similar to
a power-law fit to the ULX.  We then fitted the background-subtracted
halo spectra with a thermal model and a power-law component with
a common absorbing column and $\Gamma = 1.8$.  In the
\textit{XMM-Newton} spectra, the component is degenerate with the ULX,
and in the \textit{Chandra} spectrum the improvement to the fit is
small and the change to $Z$ is within the 90\% error bars reported in
Table~\ref{table.halo_fits}).  Thus, in contrast to \citet{kim03}, our
finding of low metallicity in the halo is not likely influenced by the
XRB population.  Fits with $\Gamma = 1.4$ (similar to the CXB)
demonstrate that the AGN contribution is also small. 

As mentioned above, the residuals in the pn and MOS fits
(Figure~\ref{figure.halo_spectra}) suggest that the \ion{O}{7} flux
near 0.58\,keV is underpredicted while the \ion{O}{8} flux is
overpredicted.  Indeed, adding a Gaussian component of zero
(instrumental) width to the \textit{XMM-Newton} spectral fits does
improve the fit (e.g. in the pn by $\Delta \chi^2 \sim 8$ when
removing a single degree of freedom for the flux of the line).  As
discussed in \citet{protassov02}, the $F$-test is inappropriate for
determining the presence of a line, but when the parameters are well
constrained (as they are in our model), the $F$-test can be
``calibrated'' using simulated spectra.  Using the \textit{fakeit}
tool in \textit{Xspec}, we simulate 100 spectra based on the best-fit
(null) model and pn response and exposure time to determine the
fraction of data sets in which the \ion{O}{7} line improves the fit by
a similar amount.  There are several cases in which this occurs, but
none have a $\Delta \chi^2$ larger than 8.  Finally, the \ion{O}{7}
complex is detected in the RGS data.

The presence of the ``line'' is unsurprising if the spectrum does
indeed require a 2-T fit.  However, it is unclear that the emission is
attributable to NGC\,891.  This is because it does not appear to be
necessary in the \textit{Chandra} spectra (although these have worse
spectral resolution) and adding the line as a Gaussian component
brings the \textit{XMM-Newton} 1-T model parameters into better
agreement with the \textit{Chandra} values
(Figure~\ref{figure.o7_contours}).  In the right-hand plot, the line
has been added and the temperatures agree.

On the other hand, from a \textit{XMM-Newton} combined image made in
the $0.5-0.6$\,keV band (Figure~\ref{figure.HI_OVII}), there is some
weak but suggestive evidence that the strong \ion{O}{7} emission is
related to the galaxy.  The emission basically tracks the inner halo
(as one expects, considering the halo continuum is bright in this
band), but there is a small extension to the northeast following the
outermost \ion{H}{1} contour.  The hole in the middle is the ULX mask.
Although the image shown is adaptively smoothed, this statement is
based on adaptive binning.  The \textit{Chandra} data do not show many
point sources missed by the \textit{XMM-Newton} mask in this region,
so the excess is likely diffuse or made up of faint point sources.

If the emission is not connected to NGC\,891, it is also not likely from
the Local Bubble or Galactic halo.  Although the Local Bubble should
radiate strongly in \ion{O}{7}, to explain the excess it would need a
flux similar to that in the 2-T model, or $3-4$\,times that predicted
by the RASS data.  Since the Galactic halo has a temperature similar
to the 1-T fit temperature ($kT \sim 0.2-0.25$\,keV), it likewise
cannot produce a strong \ion{O}{7} line without also producing a
strong \ion{O}{8} line, which is the same reason the 1-T model leaves
this excess (Figure~\ref{figure.fountain_sim}).  

It is probably not produced by solar wind charge exchange because,
while we would expect charge exchange to produce a simple line, we
expect charge exchange to produce other strong lines
(e.g. \ion{O}{8}), and \citet{carter08} find that \textit{XMM-Newton}
observations affected by charge exchange have line light curves that
are uncorrelated with continuum curves.  We find nearly perfect
correlation between the \ion{O}{7} line and nearby continuum (a
Pearson coefficient of $R = 0.997$).  There is also no record of
instrumental \ion{O}{7} lines in the XMM-ESAS manual or calibration
reports.

Overall, it is unclear whether the \ion{O}{7} flux supports or enables
the 2-T model because it seems to be peculiar to the
\textit{XMM-Newton} data and brings the 1-T temperatures into
agreement with the \textit{Chandra} data.  

\subsection{Systematic Error in $kT$ and $Z$}

As the metallicity is the crucial discriminant between the accretion
and galactic fountain scenarios, it is worth discussing the best fit
in a phenomenological sense to gauge the uncertainty in $Z$.

In principle, the temperature and metallicity can be uniquely
determined from the spectrum of a plasma in collisional ionization
equilibrium based on the lines present and continuum flux.  Once the
temperature has been established from the relative strength of
identifying lines, the metallicity can be determined by scaling to the
correct flux.  At the low spectral resolution of the CCDs, the
temperature is determined (below a few keV) by the peak wavelength of
the spectrum, since the lines produced at different temperatures are
well separated in wavelength.  The metallicity is determined in the
same way as before, but at low resolution, there are fewer patches of
(effectively) pure continuum.

\begin{figure}
\begin{center}
\includegraphics[width=1.0\linewidth]{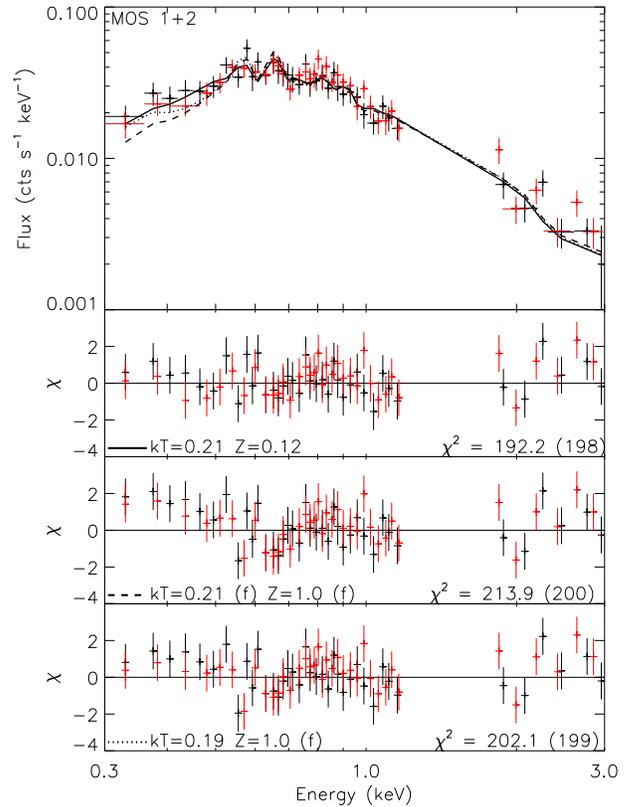}
\caption{$0.3-3$\,keV MOS 1+2 outer halo spectra with three fits overlaid.  The
  best fit (Table~\ref{table.halo_fits}) has $kT = 0.24$\,keV and $Z = 0.12
  Z_{\odot}$.  The dashed line represents the best fit with $kT$
  frozen at 0.24\,keV and the metallicity frozen at $Z_{\odot}$,
  whereas the dotted line is the best fit when $kT$ is variable ($kT =
  0.20$\,keV) but the metallicity is frozen at $Z_{\odot}$.  The
  residuals of these cases are shown below.  See text for discussion.}
\label{figure.lowz}
\end{center}
\end{figure}

The spectrum of the hot halo of NGC\,891 peaks at approximately
0.6\,keV (Figure~\ref{figure.chandra_spectra}).  The location of this
peak tightly constrains the temperature, with uncertainties in the
best-fit values of around 15\%.  The low metallicity results from the
substantial flux and shallow slope below 0.5\,keV where there are few
strong lines and the flux must be provided by the continuum.  Above
0.5\,keV there are few regions of pure continuum, so it is primarily
this region that drives the fit.  Figure~\ref{figure.lowz}
demonstrates this in the MOS\,1 and 2 spectra.  In this plot the solid
line is the best fit as reported in Table~\ref{table.halo_fits}.  The
dashed line is the best fit obtained when the temperature is frozen at
the value in Table~\ref{table.halo_fits} and the metallicity is frozen
at solar.  The primary effect of raising the metallicity is seen in
the residuals below 0.5\,keV, where the model is a poor fit.  If the
temperature is allowed to vary while the metallicity remains fixed at
solar, we can see from the dotted line in Figure~\ref{figure.lowz}
that the fit is improved but still poor below 0.5\,keV.  This effect 
would be even stronger without the ULX contribution.  Thus, our
results depend critically on the flux in this region.

There are several sources of potential bias, which can be divided into
problems with the data and problems with the modeling.  There are some
apparent differences between the \textit{XMM-Newton} and
\textit{Chandra} spectra (Figure~\ref{figure.halo_contours}), but the
differences are small in absolute terms and fall within the 3$\sigma$
statistical error contours.  Since the metallicities in all three are
small, if the calibration below 0.5\,keV is wrong, it is wrong in a
similar way for all three spectra.  The ULX is unlikely to be a source
of bias given the tight constraints on its model parameters and small
contribution below 0.5\,keV (Figure~\ref{figure.bkgfit_bkgsub}).  The
agreement in astrophysical background surface brightness and spectra
between the RASS and three instruments we use indicates that the soft
shoulder is not likely to be due to inappropriate background
subtraction, for example a higher-than-expected Local Bubble
contribution.  While fits to the instrumental backgrounds without
constraints from RASS do not constrain the Local Bubble emission well
(it drops sharply above 0.3\,keV), in a phenomenological sense it does
not matter how we fit the background as long as it is appropriately
``subtracted'' in the explicit fits, and the agreement between the
parameters (Figure~\ref{figure.bkgfit_bkgsub_contours}) indicates that
it is.

On the modeling side, one might worry about our treatment of the
photoelectric absorption, the relative abundance tables we use, the
thermal model we use, and the (non-)modeling of the possible
\ion{O}{7} excess.

Photoelectric absorption predominantly affects the softer X-rays,
thus potentially biasing $kT$ and $Z$.  In our fits to the outer halo, we fix $N_H$
at the Galactic value of $N_H \approx 6.5\times 10^{20}$\,cm$^{-2}$.
The error in this value is small compared to the effect of ignoring
the $N_H$ column in the outer halo of NGC\,891, which we do.  
Fortunately, the magnitude of the shifts, seen best in the \textit{Chandra} data, is
small: about $\pm$0.01\,keV and $\pm$0.01\,dex in $kT$ and $Z$
respectively for the 1$\sigma$ errors on $N_H$.  There appears to be
no bias towards one direction.

The {\sc wilm} abundance table is an updated version of the
\citet{anders89} table (see their paper for details).  The difference
between the tables amounts to \citep[using the][as a default]{wilms00}
a decrease in $kT$ of less than $-0.005$\,keV but a decrease in $Z$ of
$-0.04$\,dex in the MOS spectra.  Using the \citet{grevesse98} table
causes a relative increase of $+0.02$\,keV to $kT$ while decreasing
$Z$ by $-0.02$\,dex.  There is almost no difference between the
\citet{wilms00} table and the \citet{lodders03} table, which was based
on updates to solar photospheric abundances.  The primary effect of
choosing a different table is to \textit{decrease} the metallicity.

We use the {\sc apec} and {\sc vapec} thermal models with the ATOMDB
v.1.3 transition database.  Although there are differences between the
{\sc apec} and {\sc mekal} codes\footnote[1]{see
http://www.atomdb.org/Issues/mekalspex.php}, the effect on the $kT$
and $Z$ parameters in our spectra is quite small, less than
$\pm0.01$\,keV and $\pm0.02$\,dex respectively.  We note that since
the models share some calculations, this cross check does not account
for errors in some of the underlying formulas or values.  If there is
bias here, its direction is unknown.

Finally, the possible \ion{O}{7} excess, modeled as a line, does
change the temperature (Figure~\ref{figure.o7_contours}), since it is
an important diagnostic line.  However, if the excess is real it may
indicate an additional thermal component rather than a single line, so
this change in temperature should be treated with caution.
Nonetheless, the line does not appear to change the overall
metallicity much, as its presence or absence does not change the need
to model the flux below 0.5\,keV.  Adding a line component does change
the O/Fe value, counterintuitively \textit{increasing} the value
because the temperature of the 1-T model is forced higher, increasing
the emission in the Fe L-shell lines.  In the fit as reported in
Table~\ref{table.halo_fits}, the low temperature requires a relatively
higher Fe abundance to produce these lines.

Overall, the additional uncertainty incurred by our choice of
instruments, model, abundance table, and $N_H$ column do not cast
doubt on the low metallicity in the 1-T fits to the halo.  The
dominant source of uncertainty in $kT$ and $Z$ appears to be the
degeneracy between the two, which is exacerbated by the low spectral
resolution of the CCDs.  

\begin{deluxetable}{cccccc}
\tablenum{5}
\tabletypesize{\scriptsize}
\tablecaption{1-T and 2-T Derived Quantities}
\tablewidth{0pt}
\tablehead{
\colhead{Model} & \colhead{$kT$} & \colhead{$\dot{M}$} &
\colhead{$0.01-3.0$\,keV $L_X$} & \colhead{$H$} &
\colhead{$M_{\text{tot}}$} \\
& \colhead{(keV)} & \colhead{($M_{\odot}$\,yr$^{-1}$)} &
\colhead{(erg\,s$^{-1}$)} & \colhead{(kpc)} & \colhead{($M_{\odot}$)}
}
\startdata
1-T & 0.2  & 0.4     & $6\times 10^{39}$  & $5\pm2$ & $3\times 10^8$ \\
2-T & 0.1  & 3       & $4\times 10^{40}$  & $3\pm1$ & $1\times 10^8$ \\
2-T & 0.25 & $<0.05$ & $<2\times 10^{39}$ & $6\pm2$ & $<10^8$ 
\enddata
\tablecomments{\label{table.cooling} Derived quantities discussed in Section~\ref{section.discussion.cooling}.}
\end{deluxetable}

\section{Discussion}
\label{section.discussion}

\subsection{Spectral Models}
\label{section.discussion.cooling}

In the prior section we found that both the inner and outer halo are
well characterized by a 1-T thermal model with $kT \sim 0.2$\,keV and
$Z \sim 0.1 Z_{\odot}$.  Solar metallicity is strongly disfavored, as
are absorbed continuum models that do not produce emission lines.
However, the spectra also admit a physically motivated 2-T model with
solar metallicity where a cool $kT \sim 0.1$\,keV component is
dominant but a $kT \sim 0.25$\,keV component is required.  The metallicity
of the 0.1\,keV component must be close to solar (or the fit reverts to the
1-T model), but the metallicity of the 0.25\,keV component is poorly
constrained.  In our fits in Table~\ref{table.fountain_fits}, we froze it
at solar.

We now ask whether the data distinguish between these models based on
the cooling rate, scale height, and total mass involved in each model.
For the purposes of this discussion, we consider the 0.25\,keV
component in the 2-T model to be an accreted corona (possibly with
subsolar metallicity).

It is worth noting that neither model may be appropriate if the X-ray
emission is more complex.  If, for example, the X-rays originate in
cold material that is shocked by a much hotter medium \citep[as in
galactic winds;][]{strickland00b}, the 1-T and 2-T models are
insufficient to describe the true spectrum.  From a historical
standpoint, there is some reason to believe this might be true: early,
low-resolution and low-signal X-ray spectra from galactic halos were
fit by single-temperature models with low metallicity, but later fits
incorporating foreground and background components relaxed the
metallicity constraints and found a 1-T vs. 2-T ambiguity as we find
in NGC\,891.  \citet{strickland04} found that the X-ray spectra
extracted from a sample of nearby, edge-on, star-forming galaxies
could all be fit with the same 2-T model (with different
normalizations).

However, the simplest versions of the galactic fountain and IGM
accretion scenarios both predict a simple halo.  This warrants
examining the viability of these simple models in view of quantities
derived from the data.

\subsubsection{Cooling Rate}

Without heating terms, a simple estimate of $\dot{M}$ is the
luminosity divided by the specific energy, $\dot{M} = L/e$, where $e = 3kT/2\mu
m_{\text{H}}$ \citep{nulsen84} and the mean particle mass $\mu = 0.63$ for a
primordial composition.  When cooling is dominated by line emission,
as is the case for $kT \sim 0.1-0.3$\,keV, the cooling rate does not
depend strongly on density \citep{sutherland93}.  As both the 1-T and
2-T models radiate substantial energy below the 0.3\,keV cutoff the
\textit{Chandra} and \textit{XMM-Newton} CCDs, we use unabsorbed model
luminosities between $0.01-3$\,keV, summing values from the inner and
outer halo.  This is $L \sim 6\times 10^{39}$\,erg\,s$^{-1}$ in the
1-T model and $L \sim 4 \times 10^{40}$\,erg\,s$^{-1}$ in the 0.1\,keV
component of the 2-T model. For reference, the $0.3-3$\,keV model
luminosity of the halo is $L_X \approx 2\times
10^{39}$\,erg\,s$^{-1}$, which is in agreement with
\citet{strickland04}, and the $0.1-2$\,keV luminosity $L_X \sim
4\times 10^{39}$\,erg\,s$^{-1}$ in the 1-T model is in agreement with
the \textit{ROSAT} data reported in \citet{bregman94}.

The $0.01-3$\,keV luminosities yield cooling rates of $\dot{M} \sim
0.4 M_{\odot}$\,yr$^{-1}$ in the 1-T model and $\dot{M} \sim 3
M_{\odot}$\,yr$^{-1}$ in the 0.1\,keV component of the 2-T model.  The
0.25\,keV component of the 2-T model contributes $\dot{M} < 0.05
M_{\odot}$\,yr$^{-1}$.  We have corroborated these values by measuring
the density in the inner halo derived from spectra extracted near the
disk in thin strips.  We obtain the density by fitting the spectrum to
yield the model emission measure, which we use to derive the density
by assuming a volume.  The volume we use is a rectangular prism with
two axes given by the extraction box on the sky while the third axis
is estimated by assuming that the extent of the halo emission along
the line of sight is the same as the extent of the galactic disk, which
we compute as a chord across the circle at the mean galactocentric radius
of the extraction box.  Since the emission measure is proportional to
$n^2$, the error along the third axis enters as the square root of the
chord length.  In practice, the sky axes we use are 5\,kpc by 2\,kpc, 
whereas we derived densities at several radii (with imputed box depths
between $3-6$\,kpc) and found consistency within the statistical errors.
With the density and cooling function \citep[we
  use the function in][]{mathews78}, we can compute $t_{\text{cool}}$
and the total mass, thereby producing an average cooling rate.  The
values are consistent with the luminosity-based cooling rate, with
$\dot{M} \lesssim 0.5 M_{\odot}$ in the 1-T model and $\dot{M}
\sim 2 M_{\odot}$ in the 0.1\,keV component of the 2-T model
(the density tends to be overestimated based on uncertainty in the
metallicity and $N_H$, but we fix $Z = Z_{\odot}$ for the 2-T model).

These values are upper bounds if there are heating terms, in which
case $L$ becomes $L_{\text{net}}$.  The most plausible source is hot,
rising fountain material.  \citet{melioli09} find that such heating
suppresses cooling in the accreted corona entirely.  However, the
0.1\,keV fountain material is cooler than the 0.25\,keV accreted
component, besides which it is injected trans- or subsonically
\citep{houck90}, so we expect it to cool too quickly to heat the wider
corona.  In the accretion scenario, the gas is heated to the virial
temperature at the accretion shock, but in the steady state this
cannot, by definition, prevent cooling.  Thus, we consider the heating
to be a small effect.

Coronal gas will cool through lines such as \ion{O}{6}
$\lambda\lambda$1032,1038\AA, and the 0.1\,keV component of the 2-T
model is expected to produce an \ion{O}{6} $\lambda$1032\AA\
luminosity of about $L_{1032} \sim 7\times 10^{38}$\,erg\,s$^{-1}$.
The cooling rate seen in these lines may therefore favor one model.

Unfortunately, there are no \ion{O}{6} detections in NGC\,891, but
\citet{otte03} report an upper limit of $L_{\text{1032}} < 7.5\times
10^{38}$\,erg\,s$^{-1}$ assuming that the \ion{O}{6} halo is
coextensive with the X-ray halo.  Since they account only for
foreground extinction, we compute more accurate limits by using the
average \ion{H}{1} column densities at the positions of their spectra.
Assuming the typical Galactic value for reddening, $R(V) = A(V)/E(B-V)
= 3.2$, holds in NGC\,891, the maximum total neutral column seen
within the \citet{otte03} positions is $E(B-V) \sim 0.4$ very near the
disk and less than $0.1$ farther out.  Most of the flux in their
positions comes from regions with between $1-5\times
10^{20}$\,cm$^{-2}$ of intrinsic absorption.  We extrapolate the
reddening values to the UV using the curves of \citet{cardelli89}.
Using the average intensities and a column averaged over this region,
we find a limit on the unabsorbed $\lambda$1032\AA\ luminosity of
$L_{1032} < 2\times 10^{39}$\,erg\,s$^{-1}$ if the \ion{O}{6} halo is
coextensive with the X-ray halo.  \citet{edgar86} computed \ion{O}{6}
luminosities for gas cooling from above $10^6$\,K (0.1\,keV) as a
function of $\dot{M}$ for isobaric and isochoric cases.  
The \ion{O}{6} limit then corresponds to a cooling
rate of $\dot{M} < 2-3 M_{\odot}$\,yr$^{-1}$.

While this is substantially higher than the rate expected in the 1-T
model ($0.4 M_{\odot}$\,yr$^{-1}$), it is close to 
that expected in the 2-T model ($\sim 3 M_{\odot}$\,yr$^{-1}$).  If the
true value is near the upper bound, then the fountain model is
implicated, but a more sensitive observation could rule it out.  It is
also worth pointing out that the 2-T model leaves little room for a
separate warm-hot halo component.

\subsubsection{Scale Height and Masses}

The coronal scale height may also distinguish between the 1-T and 2-T
models.  Specifically, we can measure the scale height in each
component of the models assuming hydrostatic equilibrium (the
``observed'' scale height) and ask whether the value is consistent
with our expectations based on the cooling and equilibrating times.
If the observed scale height is inconsistent with these
times, the model may be unphysical.  

In hydrostatic equilibrium, $n(z) = n_0 e^{-z/H}$, where $H$
is the scale height.  Since $H = kT/\mu m_{\text{H}} g_z$ is a
function of temperature, we expect different scale heights for the 1-T
and 2-T cases.  If we assume the underlying gravitational potential is
the same in both models, then we expect the 0.2\,keV 1-T model to have
a scale height twice as large as the 0.1\,keV component of the 2-T
model.  However, the 0.25\,keV component of the 2-T model is
responsible for most of the observed emission in the \textit{Chandra}
and \textit{XMM-Newton} bandpass and would have a similar scale height
to the 0.2\,keV 1-T model.  The surface brightness profile is not
known to sufficient accuracy to distinguish between the models by
itself.  

In the 1-T interpretation, the density scale height $H$ is twice the
measured X-ray surface brightness scale height $z_0$ \citep[from this work
  or][]{bregman94}, or $H \sim 6-7$\,kpc.  For the same potential, and
ignoring potential complicating factors such as magnetic fields and
cosmic rays \citep{kalberla98}, the 0.1\,keV component of the 2-T
model would be expected to have $H \sim 3$\,kpc while the 0.25\,keV
component would have $H \sim 7$\,kpc.  Spectra extracted along and
above the disk at various positions indicate that the 1-T model
atmosphere is isothermal, so these values should be approximately
accurate.  However, to verify these values we have measured the
emission-weighted density as a function of $z$ in a set of five strips
(each 0.5\,arcmin high) above and below the disk.  We then obtain $H$
from an exponential fit to $n(z) = n_0 e^{-z/H}$.  

To maximize the signal, we fit spectra extracted from both sides of
the disk and both sides of the bulge simultaneously.  As noted above,
the surface brightness profile (Figure~\ref{figure.midplane_profile})
indicates a different $z_0$ north and south of the bulge, but as the
surface brightness is proportional to $n^2$, small differences in $H$
are exaggerated in $z_0$.  We fit the combined spectrum from each $z$
with the 1-T and 2-T models to obtain the emission measure in each
case, freezing the metallicity in the 1-T model to its overall
best-fit value and in the 2-T model to solar.  We also freeze the
absorbing column density at values estimated from the \ion{H}{1} maps
in the inner halo and at the Galactic value $N_H = 6.5\times
10^{20}$\,cm$^{-2}$ above 1\,kpc.  For the projected depth of the
extraction boxes we adopt 6\,kpc (the error incurred here goes as the
square root of the depth, and our adopted value is based on the scale
length of X-ray emission along $\hat{r}$ assuming azimuthal symmetry).

For the 1-T model, we find $H = 5\pm2$\,kpc, which is consistent with
the 7\,kpc of \citet{bregman94} as well as the 4\,kpc coronal scale
height of the Milky Way \citep{kalberla98}.  In the 2-T model we
derive $H = 6\pm2$\,kpc in the 0.25\,keV component and $H =
3\pm1$\,kpc in the 0.1\,keV component.  The uncertainties are
1$\sigma$ errors based on fitting the emission-weighted densities, but
do not include the systematic error from degeneracy in model
parameters in the spectral fitting (which is more important for the
2-T model).  For reference, we find $n_0 \sim 6\times
10^{-3}$\,cm$^{-3}$ in the 1-T model, $n_0 \sim 4\times
10^{-3}$\,cm$^{-3}$ in the 0.1\,keV component of the 2-T model, and
$n_0 \sim 2\times 10^{-3}$\,cm$^{-3}$ in the 0.25\,keV component.  The
scale heights correspond to X-ray masses (assuming the halo is
cylindrical with a radius the size of the projected optical disk
radius $R \sim 6.5$\,kpc) of $M \sim 3\times 10^8 M_{\odot}$ in the
1-T model \citep[cf. $10^8 M_{\odot}$ in][]{bregman94} and $M \sim
1\times 10^8 M_{\odot}$ in the 0.1\,keV component of the 2-T model.
The 0.25\,keV component also has $M \lesssim 10^8 M_{\odot}$.

The agreement between these values and those derived from assuming an
isothermal atmosphere with the same underlying $g_z$ above is
encouraging, but we have neglected the cooling and equilibrating
times.  We now ask whether the ``observed'' scale heights in each
model are consistent with these times. 

Hot plasma rising from the disk will reach equilibrium in about a
dynamical time.  The density and the temperature of the 0.1\,keV
component place it in the subsonic or transonic regime
\citep{houck90}, so the the sound-crossing time is a reasonable proxy
for the dynamical time (the free-fall time and the
halo-filling time are similar).  For $kT =
0.1$\,keV, $c_s \sim 0.08$\,kpc\,Myr$^{-1}$, and for $kT = 0.25$\,keV,
$c_s = 0.13$\,kpc\,Myr$^{-1}$.

0.1\,keV plasma injected at the sound speed takes about 38\,Myr to
rise 3\,kpc, during which it will cool.  The characteristic cooling
time of a plasma with a density of the value near the disk is
$t_{\text{cool}} \sim 12$\,Myr.  We can also estimate
$t_{\text{cool}}$ from the mass of the 0.1\,keV component ($10^8
M_{\odot}$) and a cooling rate of $3 M_{\odot}$\,yr$^{-1}$ from above
to find $\sim$33\,Myr.  In either case, this is less than the crossing time.
The clouds that form continue to rise to near the expected scale height before
falling back.  \citet{houck90} show that under these circumstances a
dense layer builds up.  It is possible that the surface brightness
enhancement around the disk to the north (between $1-2$\,kpc off the
disk) is this dense layer, but no corresponding layer is visible in
the \ion{H}{1} maps where the column density falls off monotonically.
Also, the north side of the bulge only has about 30\% more mass than
the south in this region.  This corresponds to less than 1\% of the
total coronal mass.

The measured scale height in the 0.1\,keV component of the 2-T model
appears to be inconsistent with what we would expect from gas actually
rising at this temperature.  A substantial amount of 0.1\,kev gas is
also required above 3\,kpc in the 2-T fit, but even if we put the
fountain model aside it is unlikely that this gas could have cooled
from the hotter medium.  This is because the 0.1\,keV material will
cool below X-ray visibility long before the 0.25\,keV gas cools to
0.1\,keV.  We might also expect to see filamentary structures in the
X-ray maps if this were the case \citep[e.g.][]{mccourt11}, but we
do not see any (Section~\ref{section.obs}). 

In contrast, the 0.2\,keV plasma of the 1-T corona moves faster and
cools more slowly and will equilibrate long before it cools.  The 1-T
model is also in better agreement with recent galactic fountain models
such as \citet{fraternali08} and \citet{marasco12} that find an
accreted corona is required to dissipate the angular momentum of
material ejected from the disk.  In these models, 10--20\% of the
total extraplanar gas mass must be accreted from the IGM.  The mass of
the \ion{H}{1} halo is $1.2\times 10^9 M_{\odot}$ \citep{oosterloo07},
so the X-ray mass in the 1-T model is $\sim$15--20\% of the total
(depending on the scale height and assumed geometry), whereas the
0.25\,keV component of the 2-T model comprises 4--8\% of the total.
The masses are unlikely to be much higher, so the 1-T model is
apparently better able to provide an angular momentum sink.  We note
that the masses of other known components of the extraplanar gas are
smaller than the \ion{H}{1} halo, with a combined total in the cold
molecular, warm neutral, and warm ionized phases of a few$\times 10^8
M_{\odot}$ \citep{garcia-burillo92,howk00}.

\begin{deluxetable}{llll}
\tablenum{6}
\tabletypesize{\scriptsize}
\tablecaption{Halo Component Masses}
\tablewidth{0pt}
\tablehead{
\colhead{Component} & \colhead{Mass} & \colhead{Depletion Rate} &
\colhead{Ref.} \\
& \colhead{($M_{\odot}$)} & \colhead{($M_{\odot}$\,yr$^{-1}$)}
}
\startdata
H$_2$        & few$\times 10^8$ & $\sim15$ & 1,2\\
\ion{H}{1}   & $1.2\times 10^9$ & $20-40$  & 3\\
\ion{H}{2}   & few$\times 10^8$ & $\sim15$ & 2\\
Warm-hot     & $1-2\times 10^8$ & $<2$     & 4, this work\\
Hot          & $2-4\times 10^8$ & $<2$     & 5, this work
\enddata
\tablerefs{(1) \citet{garcia-burillo92}, (2) \citet{howk00}, (3)
  \citet{oosterloo07}, (4) \citet{otte03}, (5) \citet{bregman94}}
\tablecomments{\label{table.component_masses}Estimated masses in halo
  components discussed in Section~\ref{section.discussion.origin}.
  The depletion rate in the cooler components is assumed to be the
  total mass divided by the free-fall timescale (based on the scale
  height) and in the hotter components it is the cooling rate.  There
  is no positive evidence for a separate warm-hot component, but we 
  cannot rule it out.}
\end{deluxetable}

\subsubsection{Concluding Remarks}

If the hot halo can be described by a simple thermal model with a
filling factor of unity, the cooling rates are on the cusp of ruling
out the 2-T model, while the measured scale height of each model is
more consistent with the 1-T model.  These arguments depend primarily
on temperature and are less sensitive to the details of the physical
model.  For example, 0.2\,keV plasma in a 1-T fountain model would
still equilibrate before cooling.  Thus, we conclude that the 1-T low
metallicity model is better at explaining the existing data than the
2-T solar metallicity model, but these considerations do not prove an
origin.  The low metallicity in the 1-T model does not appear to be
due to incorrect subtraction of the X-ray background or other
systematic effects, but it is possible that the spectrum is
undermodeled \citep[cf.][]{strickland00}.  The 400\,ks simulated
\textit{XMM-Newton} spectrum in Figure~\ref{figure.fountain_sim} would
rule out a 1-T model based on Fe-L lines.  If the 2-T model were still
disfavored, such results would point toward a more complex hot halo.
It is also possible that the 2-T model represents a fountain$+$accretion
scenario in which the hotter component (whose metallicity is poorly 
constrained, but somewhat above the 0.1$Z_{\odot}$ of the 1-T model)
represents accreted gas that has equilibrated at a different scale
height than the cooler fountain gas.  The simulated \textit{XMM-Newton}
spectrum in Figure~\ref{figure.fountain_sim} would strongly constrain
the metallicity in the 2-T model if the 1-T model has already been ruled 
out.  In this case, the accretion component would dominate farther
from the galaxy.  

\subsection{Origin of the Extraplanar Gas}
\label{section.discussion.origin}

We now turn to the origin of the halo gas.  In the classic galactic
fountain scenario, superbubbles break out of the disk, venting hot gas
to the halo which cools and falls back as \ion{H}{1} clouds.  In this
section, we will first demonstrate that this model cannot explain the
extraplanar gas in NGC\,891 if it is in the steady state.  We will
then consider alternatives, including cold accretion and prior
outflows, before placing NGC\,891 in the context of other massive
late-type galaxies.  We conclude by offering a speculative picture
that NGC\,891 and several other galaxies represent a period of
enhanced star formation in the disk following a nuclear starburst that
has since wound down.

\subsubsection{Galactic Fountains and the Steady State}

Considering the relevant timescales and disparate masses in the
components of the extraplanar gas
(Table~\ref{table.component_masses}), it is evident that the atomic
component could not have cooled from the hot corona in the steady
state.  This is because the cold ($T < 10^4$\,K) gas free falls back
onto the disk (in $t_{\text{ff}} \sim 30-50$\,Myr) at a rate of
$\dot{M} \sim 30 M_{\odot}$\,yr$^{-1}$, whereas the total cooling rate
from the X-ray and UV emitting gas is $\dot{M} < 3
M_{\odot}$\,yr$^{-1}$.  This rate could be as low as $\dot{M} \lesssim
0.5 M_{\odot}$\,yr$^{-1}$ in the 1-T accretion model if the only source
of \ion{O}{6} luminosity is cooling coronal gas.  In this case, the
instantaneous gas consumption rate by star formation \citep{popescu04} 
is more than 4\,times higher than the accretion rate. 

The halo components could exist in a steady state if the \ion{H}{1} is not
supplied by the hot gas, but rather rises into the halo via a
(primarily cold) galactic fountain \citep[e.g.][and
variants]{fraternali06}.  The viability of a cold fountain depends on
its ability to expel enough gas and to keep it cold.  In our view, if
the fountain is powered by superbubble breakouts (the most natural way
to launch material more than 1\,kpc above the disk), it fails to meet
both criteria as we argue below.  This rules out a halo formed by the
steady-state action of a galactic fountain, hot or cold. 

First, it is worth laying out our superbubble breakout paradigm.
Superbubbles form around OB associations as stellar winds and SNe
evacuate a region in the surrounding ISM.  Once a bubble has formed,
its structure is thought to be similar to the picture given in
\citet{weaver77}, \citet{smith01b}, and \citet{maclow88}.  Near the
core, the hypersonic wind or supernova blast wave freely expands.
Farther out, the flow encounters a stagnation shock and heats to X-ray
emitting temperatures ($10^6 < T < 10^8$\,K).  This hot medium fills
most of the bubble.  SNe that explode after the bubble has formed do
not form remnants like isolated SNe, but instead convert their kinetic
energy into thermal energy by heating the already hot interior via
weak shocks.  Radiative losses from the very hot material are small.
The bubble is bounded by a shell of swept-up mass, which is thin early
on but thickens in developed bubbles.  The hot gas within evaporates
mass from the inner layer of this shell via thermal conduction, and in
a developed bubble the hot mass is dominated by evaporated gas.  The
shell itself is not significantly depleted by this process. 

Once a bubble's diameter exceeds about three times the scale height of
the disk \citep{maclow89}, it will break out into the halo.  The
breakout is characterized by accelerated upward shell expansion, which
leads to Rayleigh-Taylor instabilities that disrupt the shell and
allow the hot gas to vent.  However, less than 10\% of the shell mass
erupts into the halo \citep{maclow89}, and this rises into the halo as
fragmented clouds.  The hot gas evacuates as a pressure-driven
outflow.  This scenario finds some support in NGC\,891: in a study of
the emission from dust and diffuse H$\alpha$ above the disk,
\citet{rossa04} do not find large, organized ``chimneys''.  Rather,
the evidence indicates that the bubbles vent through a disorganized
medium. 

To determine whether the outflow is predominantly hot or cold requires
estimating the masses and energies involved.  We do not have direct
observations in NGC\,891 (although dust filaments indicating bubbles
and a somewhat porous disk are present), but we can estimate the cold mass from the
\citet{boomsma08} study of \ion{H}{1} holes in NGC\,6946, a nearby
face-on spiral with a similar mass and SFR as NGC\,891.  We also consult
values derived from other face-on spirals that are not as apparently
analogous to NGC\,891.  The hot mass
is uncertain, but we can constrain it by considerations from nearby
superbubbles that have been studied in the X-rays.  The fiducial
values we derive are given in Table~\ref{table.fiducial}.

We assume that the cold mass ejected into the halo is 10\% of the
mass of the swept-up shell at the time of breakout \citep{maclow89},
although it may be less.  As will be clear later, it is difficult
from an energetics perspective to expect much more.  The breakout
size of the hole depends on the scale height of the disk, which in
NGC\,891 is about 200--400\,pc \citep{xilouris98}.  We adopt 300\,pc
as a fiducial value, meaning a bubble at breakout has a radius
of about 450\,pc, but we note that the scale height varies with
distance from the galactic center.   
\citet{boomsma08} estimated shell masses in
deep observations of \ion{H}{1} holes in NGC\,6946 by measuring the
ambient column density around each hole and the area of the hole.
Because they measured $N_H$ in regions including the swept-up shell
itself, the resulting mass is an upper bound.  They find an
exponential distribution of hole size with radius and an average
missing mass per hole of $M < 10^7 M_{\odot}$.  This is consistent
with the swept-up shell mass in the \citet{maclow89} simulations of $M
\sim 3\times 10^6 M_{\odot}$ at breakout, and the \citet{boomsma08}
value is biased towards higher mass by the regions used to measure
$N_H$, the angular resolution (their synthesized beamsize is 390\,pc,
which is close to the breakout size), and hole identification
algorithms (since $M \propto N_H r^2$, larger holes have much more
mass).  For example, \citet{boomsma08} identify 121 holes in NGC\,6946
while \citet{bagetakos11} find 56 in the same galaxy.  

We can also compare these results to the face-on galaxies M101 \citep{kamphuis93}, 
M31 \citep{brinks86}, and NGC\,2403 \citep{thilker98} where similar measurements
have been made at high resolution and sensitivity.  M101 has a similar
mass and SFR to NGC\,891 \citep[$\sim 5 M_{\odot}$\,yr$^{-1}$ in][]{kuntz10}, 
whereas M31 has a similar mass but much lower
SFR, and NGC\,2403 is about ten times less massive but has a relatively
high SFR for its mass \citep[$\sim 0.6 M_{\odot}$\,yr$^{-1}$ in][see erratum]{heald11} 
and a lagging \ion{H}{1} halo, suggesting it is a lower mass analog to NGC\,891.  
The results suggest that the amount of mass swept up for a shell of a
given size and age is roughly the same across the galaxies.  
\citet{kamphuis93} find relatively large holes in M101
comparable to or larger than the supershells seen in NGC\,6946, with
similar masses and dynamical ages.  On the other hand, \citet{brinks86}
generally find smaller, younger holes in M31 than NGC\,6946, with a typical
missing mass of only $10^5 M_{\odot}$ per hole for holes with radii of about
100\,pc and kinematic ages of 2--30\,Myr.  
In NGC\,2403, \citet{thilker98} identify both
kinematically young (expanding) shells and older stalled shells with
a missing mass range of $10^6-10^7 M_{\odot}$ per hole and ages between
1--200\,Myr.  The range of bubble sizes, ages, and masses generally
supports the \citet{maclow89} picture and indicates that bubbles large
enough to break out of the disk in NGC\,891 would
have masses between $10^6-10^7 M_{\odot}$.  Based on the distribution of
the hole sizes and masses in the four face-on galaxies we consider, we
adopt a fiducial missing mass of a $r = 450$\,pc hole in NGC\,891 to be
$M = 3\times 10^6 M_{\odot}$.  The cold mass expelled upwards at breakout
would then be $M \lesssim 3\times 10^5 M_{\odot}$ per hole \citep{maclow89}.
\citet{maclow89} find that the ejected shell material has upward velocities
between 50--100\,km\,s$^{-1}$, so the kinetic energy required is 
$7\times 10^{51}$\,erg$< E < 3\times 10^{52}$\,erg.
This is comparable to the total kinetic energy in the rest of the shell,
but small compared to the thermal energy stored in the hot gas that drives
the outflow (see below).

The hot mass in bubbles is quite uncertain, but we estimate the mass
by assuming a spherical bubble just prior to breakout whose size is 3
times the scale height of the disk in NGC\,891.  A bubble filled with
hot gas has a mass $M/f \sim \tfrac{4}{3} \pi r^3 \mu m_{\text{H}}
n_e$, where $f$ is the filling factor.  Although we expect the bubble
to be filled with hot material (the sound speed of the hot gas is much
faster than the bubble expansion rate), much of the (soft) X-ray emission
comes from the inner edge of the shell where supernova blast waves
impact it \citep{chu95,jaskot11}.  The rest of the bubble is
presumably filled with even hotter, more tenuous gas.  Thus, the
emission measure of the hot component that dominates the mass is
$n_e^2 f V$.  Electron densities measured in X-ray observations of
nearby superbubbles range from $n_e \sqrt{f} \sim 1-100\times
10^{-3}$\,cm$^{-3}$ for bubbles between $100-600$\,pc across
\citep[e.g.][]{cash80,points00,dunne01,smith04}.  For the breakout bubble
radius of 450\,pc, the mass is $M/f \sim 5\times 10^5 M_{\odot}
h_{\text{kpc}}^3 n_{-3}$, where $h_{\text{kpc}}$ is the disk scale height
in kpc and $n_{-3}$ is in units of $10^{-3}$\,cm$^{-3}$.  For fiducial
values of $h = 0.3$\,kpc and $n = 0.01$\,cm$^{-3}$, we find $M/f \sim
10^5 M_{\odot}$.  

\begin{deluxetable}{ll}
\tablenum{7}
\tabletypesize{\scriptsize}
\tablecaption{Superbubble Fiducial Values at Breakout}
\tablewidth{0pt}
\tablehead{
\colhead{Quantity} & \colhead{Value}
}
\startdata
SNe before breakout& $\sim$100\\
Energy budget      & $\sim10^{53}$\,erg\\
Shell mass         & $3\times 10^6 M_{\odot}$ \\
Disk scale height  & 300 pc\\
Radius at breakout & 450 pc\\
Breakout cold mass & $3\times 10^5 M_{\odot}$ \\
Breakout hot mass  & $5\times 10^4 M_{\odot}$ \\
$n_e$              & 0.01 cm$^{-3}$\\
$T_e$              & $4\times 10^6$ K
\enddata
\tablecomments{\label{table.fiducial}Fiducial values used in
  Section~\ref{section.discussion.origin}.  We consider these
  reasonable estimates of the true values, which are hard to
  determine.  See text for details.}
\end{deluxetable}

There is considerable uncertainty regarding the filling factor of the
X-ray emitting material in nearby superbubbles, and we do not have
direct information from NGC\,891.  However, we suggest that a reasonable
lower bound for the mass of hot material in a bubble at breakout is 
$M \sim 10^4 M_{\odot}$, which we justify presently.
Over 100 SNe are thought to be required for a bubble to break out of the
disk of a galaxy like the Milky Way \citep[e.g.][]{maclow89,melioli08},
and by extension NGC\,891.  This will release about $10^{53}$\,erg (stellar
winds will also contribute to the kinetic luminosity).  Of this, a
few$\times 10^{52}$\,erg is contained in the average shell described in
\citet{boomsma08}.  The majority of the energy is either radiated or
goes into heating gas evaporated from the inner wall of the shell to
X-ray emitting temperatures \citep{weaver77}.  Nearby superbubbles have
interior temperatures ranging from $1-8\times 10^6$\,K \citep[e.g.][]{cash80,dunne01,smith04},
so as a fiducial temperature we take $T = 4\times 10^6$\,K.  The specific
energy of material at this temperature is $e = 8\times 10^{14}$\,erg\,g$^{-1}$,
or $e = 1.6\times 10^{48}$\,erg\,$M_{\odot}^{-1}$.  If about 30\% of the
superbubble energy budget goes into heating evaporated gas, about
$2\times 10^4 M_{\odot}$ of gas could be heated to the fiducial temperature.
The actual value may be higher depending on radiative losses.  

Extrapolations from observed superbubbles also support $M \sim 10^4 M_{\odot}$ of hot
gas as a lower limit in large bubbles.  If the hot masses inferred in 
smaller bubbles \citep[e.g.][]{dunne01} are scaled up to bubbles of $r = 450$\,pc,
the masses are $M \sim 1-10\times 10^4 M_{\odot}$, and \citet{yukita10}
find in spectral fits to \ion{H}{2} regions in NGC\,2403 that the ratio of hot gas
to swept-up mass is 6--10\% (for our fiducial shell mass of $M = 3\times 10^6 M_{odot}$,
this is $M \sim 18-30\times 10^4 M_{\odot}$ of hot material, assuming this ratio
holds throughout the lifetime of a massive bubble).  

These considerations suggest a hot mass somewhere between $1-10\times
10^4 M_{\odot}$, so for the sake of argument we take as a fiducial
value the hot mass in a $r=450$\,pc bubble to be $M = 5\times 10^4
M_{\odot}$.  This would give a filling factor $f = 0.5$ in the bubble
described above, but we emphasize that the hot mass and filling factor
of the soft X-ray emission is poorly known in many bubbles. 

Thus, at breakout our fiducial bubble (Table~\ref{table.fiducial}) launches
$3\times 10^5 M_{\odot}$ of cold shell material towards the halo and contains
$5\times 10^4 M_{\odot}$ of hot ($T = 4\times 10^6$\,K) 
material whose thermal energy ($E = 8\times 10^{52}$\,erg) is similar to
or greater than the kinetic energy in the entire shell.  The erupting shell
material is therefore driven rapidly upwards even though it dominates the mass.
Since the hot gas flows faster than the ejected shell fragments, it can 
evaporate the neutral material just as in the confined bubble. 
At our fiducial temperature of $4\times 10^6$\,K, the sound speed is $c_s \approx
150$\,km\,s$^{-1}$, whereas the shell ejecta travel at
$50-100$\,km\,s$^{-1}$ in the simulations of \citet{maclow89}.  Thus,
the cool parcels can be considered as islands in a stream.  The thermal
energy of the hot gas is sufficient to raise about $2\times 10^5 M_{\odot}$
of cold material to $10^6$\,K, so it is energetically feasible for the 
hot gas to become the dominant component (there is more than enough energy
to merely evaporate the cold material by heating it to, say, $10^5$\,K).

Whether the hot component becomes dominant is unclear, 
since evaporation occurs via thermal conduction in a boundary layer around 
the fragments and most of the hot material would not participate.   On
the other hand, the fragments are rising into the hot bath of the extant
halo.  A significant unknown is the effect of magnetic
fields, which could suppress conduction relative to the
\citet{spitzer62} value if the fragments are wrapped in field lines,
which are seen to be expelled along with the gas \citep{heald12} in
at least one superbubble in NGC\,6946.  However,
turbulence could mitigate this effect \citep[cf. galaxy clusters
  in][]{narayan01}.  In the bubble described by \citet{heald12}, both the large-scale
fields around the bubble and turbulent fields entrained in the hot plasma
are uplifted.  The speed of the hot gas flow is also important,
since gas that is too fast shocks the cooler entrained gas instead of
evaporating it \citep{strickland00}.  Pure hydrodynamic models suggest
that virtually all of the gas rises hot \citep{melioli08}, but the
detection of polycyclic aromatic hydrocarbons (PAHs) to a scale height
of $\sim$500\,pc in NGC\,891 \citep{rand11} may indicate that a
substantial amount of material rises cold.  The difference between PAH
features in the disk and halo suggests that the cool fountain gas does
experience shocks as it rises, but that these do not heat the gas to
X-ray emitting temperatures (it is also possible the grain destruction
time exceeds the flow time).  

The proportion of cold fragments to hot gas will decline with distance
above the disk since some of the cold material that is not evaporated
will fall back below 1\,kpc (above which it would unambiguously be
considered halo material).  \citet{maclow89} find that the fragments
have upward velocities between $50-100$\,km\,s$^{-1}$, and on the
lower end these fragments will reach an apex below 1\,kpc.  

Considering these factors, we suppose that less than half of the
fountain material that makes it into the halo is cold.  This amounts
to a few$\times 10^5 M_{\odot}$ of cold material injected into the
halo per hole and a comparable mass of hot gas.  Thus, a galactic
fountain powered by supernova feedback seems unlikely to be truly
``cold''.  

More importantly, it is evident from these values that the proportion
of cold and hot material in the fountain is a moot point in NGC\,891
because the fountain envisioned above does not expel enough mass to 
sustain the halo regardless of the temperature.  This can also be seen
in NGC\,6946, whose \ion{H}{1} holes we used to derive our values.
For our fiducial values, the
total amount of gas that rises into the halo in NGC\,6946 is no more
than $\sim$15\% of the total extraplanar \ion{H}{1} \citep{boomsma08}
and coronal gas \citep{schlegel03}.  Even with optimistic assumptions
about the true hole size and amount of kinetic energy involved, it is
difficult to envision increasing the outflow mass by an order of
magnitude without ejecting the entire swept-up shell mass of each
contributing hole, which is impossible for isolated holes.  The
situation is even worse in NGC\,891, which has a similar SFR to
NGC\,6946 (and therefore presumably about the same number of holes)
but four times more extraplanar \ion{H}{1}. 

This problem is circumvented if the extraplanar gas spends more time in
the halo than the lifetime of the progenitor bubble. 
The lifetime of the \ion{H}{1}
holes is limited by the slowing expansion velocity (once it drops
below the characteristic velocity in the surrounding gas, the bubble
breaks up) and interactions with neighboring bubbles (converging flows
from two expanding shells can promote star formation which will disrupt
that part of the shell), and is perhaps 80\,Myr \citep{boomsma08}.  
Hot gas may reside in the halo longer than this, but we have already
established that the \ion{H}{1} halo cannot have cooled from the 
X-ray emitting gas if it is in the steady state.  The time spent in 
the halo by the cold gas is about twice the free-fall time, or $30-80$\,Myr for the ejection
speeds in \citet{maclow89}.  Since the expulsion of cold mass from the bubble
is a one-time event occurring at breakout, we expect the cold material
launched into the halo to return to the disk within a bubble lifetime
\citep[although it will not refill the hole since the galaxy rotates and the
fountain material has angular momentum;][]{melioli08}.  Additional hot material
may leak out of the hole following the breakout, but for an isolated hole this
will be subsequent stellar wind and SNe material.  The masses of these 
components is quite small; even if we assume that the hot gas does not
escape the blown-out bubble before evaporating additional material from the
shell, the evaporated component described above was a small fraction of the
total swept-up shell mass.  

Thus, neither a classic nor a cold fountain seem capable of producing
the extraplanar \ion{H}{1} in NGC\,891 in the steady state at the
current SFR.  Although our argument for NGC\,891 is largely
circumstantial, we note that NGC\,6946 has a similar mass, SFR, and is
disk-dominated like NGC\,891 with a comparable disk scale height.
Since stellar feedback must be tied closely to the SFR, there is no
reason to expect a much higher hole rate in NGC\,891.

It is not clear if this imbalance occurs more generally.  There are
not many non-starburst galaxies with the X-ray, UV, H$\alpha$,
\ion{H}{1}, and radio continuum data required to establish an
\ion{H}{1} mass and cooling rate (starbursts are not expected to be in
equilibrium).  This is largely due to the paucity of X-ray and
\ion{O}{6} measurements around non-starburst massive spirals
\citep[see][]{tyler04}.  However, among galaxies with relatively
complete coverage, there are others that appear to fall into this
category.  As we have established, the \ion{H}{1} halo of NGC\,6946
cannot have been ejected by the observed holes unless the expelled
mass rate is an order of magnitude larger, but it also cannot have
cooled from the hot corona \citep[which has a bolometric $L_X <
10^{39}$\,erg\,s$^{-1}$;][]{schlegel03}.  M101 likewise has a 
similar mass and SFR with a bolometric halo X-ray luminosity of
$L_X < 10^{39}$\,erg\,s$^{-1}$ \citep{kuntz10} with \ion{H}{1} holes
detected throughout the disk \citep{kamphuis93}.  The halo X-ray cooling
rates implied are $\dot{M} < 0.05 M_{\odot}$\,yr$^{-1}$.  NGC\,2403
has the same issue: the X-ray emitting mass conceivably in the halo is
only 10\% of the extraplanar \ion{H}{1} \citep{yukita10}. 

Unfortunately, the picture is not clear in normal edge-on spirals.
NGC\,891 has the highest X-ray surface brightness of a non-starburst
edge-on galaxy \citep{bregman97}, so much of the X-ray work on edge-on
spirals has focused on starbursts.  A few exceptions include NGC\,4631
\citep{wang95,strickland04} and NGC\,4565 \citep{vogler95}, but both are
interacting with companions, and \citet{strickland04} argue that
NGC\,4631 is experiencing a weak starburst.  NGC\,4565 has a similar
mass to NGC\,891, but a much smaller SFR \citep[$\sim0.7
M_{\odot}$\,yr$^{-1}$ in][see erratum]{heald11} and evidence only for a weak
\ion{H}{1} halo \citep{zschaechner11}.  NGC\,4631 has a large halo,
but \citet{rand94} argue based on kinematics that most of the
\ion{H}{1} far from the disk is due in part to the not perfectly
edge-on inclination (85$^{\circ}$), tidal interaction, and a large
scale height in the outer part of the galaxy.  \citet{strickland04}
included two non-starburst edge-on galaxies in their \textit{Chandra}
sample aside from NGC\,891 (NGC\,6503 and NGC\,4244), but both are
substantially less massive than NGC\,891 and significant X-ray
emission was detected in neither.  Finally, \citet{li07} detected
diffuse X-ray emission in M104 with a bolometric X-ray luminosity of a
few$\times 10^{39}$\,erg\,s$^{-1}$, but M104 is also at least several
times more massive than NGC\,891 and \citet{li07} argue that the halo
is X-ray under-luminous while \citet{bajaja84} find a relatively small
amount of \ion{H}{1} in the galaxy.  

If the halo is not in the steady state, there are two basic
possibilities for the origin of the extraplanar gas: cold accretion
and prior outflows.  

\subsubsection{Cold Accretion}

Cold accretion (including the recycling of gas ripped out of a disk by
galactic interaction) can occur via the infall of intergalactic
\ion{H}{1} clouds or through tidal stripping and cannibalization of
small satellite galaxies.  The former is disfavored by the paucity of
intergalactic \ion{H}{1} \citep[see discussion in][]{sancisi08} and
the small amount of counter-rotating extraplanar \ion{H}{1} in
NGC\,891 \citep{oosterloo07}.  Recent simulations also indicate that
accretion of fresh material primarily occurs in the hot phase
\citep{oppenheimer10,joung12}.  While it is possible that a large
reservoir of photoionized hydrogen resides outside the visible
\ion{H}{1} halo, this would require an extremely extended halo.  The
\citet{oosterloo07} contour map extends down to $10^{19}$\,cm$^{-2}$
and monotonically declines from the center.  The scale height suggests
that the column drops below $10^{18}$\,cm$^{-2}$ within several kpc of
the last contour.  At this point, we would expect the \ion{H}{1} to be
photoionized.  Although the column density likely continues to drop
beyond this, if we assume that the space around NGC\,891 is filled
with a uniform column of $10^{18}$\,cm$^{-2}$ ($10^4
M_{\odot}$\,kpc$^{-2}$), it would take about $10^5$\,kpc$^{2}$ to make
up $10^9 M_{\odot}$ of ionized hydrogen.  If we further assume that
the photoionized halo is circular in projection and centered on
NGC\,891, the radius required is about 200\,kpc.  Replenishing the
visible \ion{H}{1} halo from such a reservoir cannot explain the
observed mass, and it is not clear that such a halo would rotate with the 
galaxy to such distances.

A more plausible scenario is that the halo is made up of tidally
stripped gas from a satellite galaxy and gas ripped out of the disk of
NGC\,891 by the same interaction.  However, there is no direct
evidence of an interaction such as tidal streams, stellar shells, or a
warped disk that would be particularly visible in an edge-on galaxy.
The absence of these features and the kinematics of the extraplanar
\ion{H}{1} indicate that the time elapsed since the posited
interaction is a few times the galactic rotation timescale of a
few$\times 10^8$\,yr.  In this case, however, we would expect gas
within a few tens of kpc to have fallen back onto the disk.  There are
a few candidates for recent interactions.  First, the companion galaxy
UGC\,1807 is a small but gas-rich galaxy with about 10\% of the mass
of NGC\,891 and is sufficiently far away to account for the lack of
tidal features \citep{oosterloo07}.  However, \citet{oosterloo07} find
that the amount of accretion that can be explained by this interaction
is only $\dot{M} \gtrsim 0.1 M_{\odot}$\,yr$^{-1}$.  

Two other signatures of recent or ongoing accretion are described in 
\citet{mouhcine10}, namely an extended (10\,kpc above the disk), flat
(disk-like) envelope of halo stars and a giant rosette pattern of tidal
streams extending out to 30\,kpc from NGC\,891.  We do not believe the
former can account for much recent cold accretion because such extended
``disks'' form on long timescales of a few Gyr or more when dwarf galaxies
are cannibalized \citep{penarrubia06}.  Moreover, \citet{mouhcine10} also
find a similar extended halo around NGC\,2683, a highly inclined ($80^{\circ}$)
galaxy with a similar total mass to NGC\,891 but no substantial \ion{H}{1}
halo \citep{kennedy09}.  Such envelopes may be common, but evidently
need not correlate with recent cold accretion.  Regarding the rosette
tails, \citet{mouhcine10} argue that they likely originate from a single
accretion event based on the similar morphology to disrupted dwarf galaxies
in $N$-body simulations.  If the galaxy were gas-rich, it could provide a
substantial cold accretion rate over a relatively long period of time,
but \citet{mouhcine10} argue based on stellar ages (no stars younger than
a few Gyr were seen in the streams) that the companion had little \ion{H}{1}
compared to UGC\,1807.  Although the instantaneous cold accretion rate could
be quite high upon absorption of a particular stream, we would expect an
average accretion rate from a gas-poor dwarf with a pre-merger gas mass 
$M_{\text{HI}} \sim 10^8 M_{\odot}$ to be no more than 
$\dot{M} \sim 0.1-0.2 M_{\odot}$\,yr$^{-1}$, considering that the lifetime 
of the stellar streams is between $0.5-1$\,Gyr \citep[e.g.][]{binney08}.  
This is similar to the imputed contribution from UGC\,1807 of 
$\dot{M} \gtrsim 0.1 M_{\odot}$\,yr$^{-1}$ \citep{oosterloo07}.  

\subsubsection{Prior Outflow}

The remaining possibility we consider is a prior hot outflow that cooled
to form the \ion{H}{1} halo.  There are two basic models, including a 
vigorous fountain (powered by higher star formation distributed
throughout the disk) or a nuclear wind.  Plausible cooling mechanisms
that would produce the observed halo can be found for either case, but
as winds are ubiquitous in starburst galaxies \citep{veilleux05}, it may
be that a powerful fountain is always accompanied by a nuclear wind.

Typical outflow rates in winds appear to be $\sim2-3$ times the
instantaneous SFR \citep[e.g.][]{bouche12}, although it is unclear how
much mass they carry because the mass carried in each component is
hard to measure \citep[see][]{strickland09}.  If material flowed out of
the disk at $30 M_{\odot}$\,yr$^{-1}$, this would imply a SFR of
$10-15 M_{\odot}$\,yr$^{-1}$ during the active phase, which is 
plausible for massive starbursts.

As the putative outflow must have occurred long enough ago for the \ion{H}{1}
to have cooled from the hot phase, a variety of
parameters are possible.  The minimum age of the outflow can be
constrained by the cooling time of the gas and (in the case of the
wind) the fallback time from large distances.  In either case, 
ages of 50\,Myr or more are expected.  The absence of wind signatures
\citep[cf. edge-on starburst galaxies in][]{strickland04} and the duration
of starburst episodes place lower limits consistent with this value.
Thus, it is neither possible to rule this out nor to make verifiable
predictions, so what follows is speculation.

NGC\,891 belongs to a class of massive, disk-dominated, late-type
galaxies with relatively high disk-distributed SFR that are not
classified as starbursts.  Nearby examples include NGC\,6946,
NGC\,2403, M101, NGC\,4631 \citep[although it is classified as a
starburst by][]{strickland04}, and NGC\,5775 \citep[classified as a
starburst in][]{li08}.  Of these, NGC\,4631 and NGC\,5775 show some
evidence for a wind-like nuclear outflow in the X-rays, with the case
being stronger in NGC\,4631.  As one might expect, such galaxies tend
to have more prominent halos than other massive spirals with lower SFR
\citep[e.g.][]{rand96,dahlem97}, with an active fountain implicated as
a result.  The existence of \ion{H}{1} halos that appear too massive
to have cooled in the steady state in NGC\,891, NGC\,6946, M101, and
NGC\,2403 (more precisely, the X-ray cooling rate and expected cold
outflow appears unable to replace gas lost via free-fall) 
suggests that NGC\,891 may simply be an extreme case of a
more frequent phenomenon, and that prior outflows may also be
necessary in these galaxies.  The case of NGC\,2403 is especially
interesting in this vein because a \textit{Chandra} observation shows
X-ray emission that is more centrally concentrated than the
star-forming regions \citep{yukita10}.  Since NGC\,2403 has a
relatively high SFR ($\sim 1 M_{\odot}$\,yr$^{-1}$) for its mass
($M_T \sim 10^{10} M_{\odot}$) and the X-ray emission (if produced by star
formation) represents a significant amount of mass, NGC\,2403 may have
had a nuclear starburst that has propagated outwards or shut off while
the disk star formation continued \citep{yukita10}.  Another hint of a
prior outflow is found in M104, which has a curiously low SFR for its
mass, where \citet{li07} argue that a nuclear wind (in their model
powered by Type 1a SNe) has removed much of the material in the inner
region of the galaxy, an argument supported by the absence of
\ion{H}{1} in the same region \citep{bajaja84}.

We therefore speculate that the high SFR is the remainder of a
starburst phase that ejected a substantial amount of material or else
fresh star formation fueled by the return of this material to the
disk.  This scenario is attractive
because it explains the evidence that points towards a
vigorous ongoing disk--halo interaction (summarized in
Section~\ref{section.introduction} for NGC\,891), the apparent
disconnect between the \ion{H}{1} and X-ray halos (and perhaps the low
metallicity of the latter, if it is accreted), and why starbursts
generally do not have extended hydrostatic X-ray halos or prominent
extraplanar \ion{H}{1} (although the latter may be due to
photoionization of extant hydrogen).  This appears to be true in
NGC\,253, NGC\,3628, NGC\,3079, and the less massive NGC\,4945,
NGC\,1482, and M82, where outflow morphology is seen in the X-rays
\citep{strickland04} and other bands, but only modest signs of
extraplanar \ion{H}{1} \citep{dahlem97}.  The outflow rates required
to explain the \ion{H}{1} halos are easily met in starbursts.  On the
other hand, if the \ion{H}{1} halos develop as hot gas cools, we would
expect to see X-ray luminous post-starburst galaxies.  This seems
likely even if much of the \ion{H}{1} halo is formed by recombination
of extant photoionized gas in the wake of the starburst \citep[cf. the
starburst galaxy NGC\,253 where the \ion{H}{1} disk is smaller than
the optical disk;][]{boomsma05}.  Although the sample of nearby
massive spirals with X-ray observations is small, NGC\,891 is the most
X-ray luminous of the nearby non-starburst spiral galaxies.  

In principle, this picture is consistent with either the 1-T or 
2-T coronal models depending on the type of outflow.  It is possible
that the observed corona is dominated by accreted material that has
fallen in during and since the formation of the \ion{H}{1} halo.  
Such a situation could arise in an outflow scenario if a distributed
starburst (i.e., non-nuclear) ``fountain'' expelled a large amount of 
material ($\sim 10^9 M_{\odot}$) in a relatively short period of time 
(a few tens of Myr).  If the material was expelled via superbubble 
breakouts at a constant rate and entered the halo at several million 
degrees, the initial material would equilibrate before the end of the
starburst and begin to cool.  Although its cooling time would be quite
high at first, the scale height of the corona would be determined by
the temperature, so additional fountain action would increase the 
coronal density (and luminosity), thereby decreasing the cooling time.  
As more material accumulated in the halo, the cooling time would 
eventually shrink until the cooling rate was comparable to the 
fallback time of cold material, and the coronal gas would rapidly
cool to form a large \ion{H}{1} halo.  If there were steady accretion
from the IGM throughout this period, the hotter material further out
would survive the rapid cooling phase, leaving the halo possibly 
dominated by accreted gas.  Whether this would actually happen depends
on the fountain ejection rate, the temperature, and the details of the
cooling.  For example, \ion{H}{1} clouds may primarily form via the
thermal instability, and the putative ``cooling flow'' would form the
most cold gas nearest the disk, so the \ion{H}{1} halo might not extend
more than a few kpc from the disk.  More importantly, it is not clear
that such powerful distributed ``starbursts'' occur.  On the other hand,
it is easier to explain such an outflow in the 2-T model of the halo
(if the 0.25\,keV component has near solar metallicity and the 0.1\,keV
component represents the present-day fountain).

Overall, the hypothesis of a previous outflow is severely hampered
by the dearth of meaningful constraints, but the present-day fountain
and cold accretion do not seem sufficient to explain the massive
\ion{H}{1} halo in a steady state. 

\subsubsection{Concluding Remarks}

We have argued that the massive \ion{H}{1} halo in NGC\,891 (and, we
speculate, in a handful of other nearby galaxies with sufficient data)
cannot be supported by the instantaneous SFR or cooling from the X-ray
halo.  However, a rigorous
examination of the fountain models and halo masses must be made in
order to confirm this picture.  For example, it is possible that the
interaction between multiple breakouts within a small patch of the
disk \citep{melioli09} might conspire to eject much more mass, or that
new bubbles forming below the falling back fragments of old ones can
lift much more cold material than the breakout of a bubble from a
pristine disk.  Our arguments are based on the latter scenario as it
appears in maps of \ion{H}{1} holes that the overlap fraction is
small, but this may be a bias of the hole detection algorithm.  An
additional and important unknown is the amount of warm-hot material in
the halos of various galaxies, although we would expect a separate
warm-hot component to equilibrate with a small scale height.  In NGC\,891,
at least, such a component could not explain the substantial extraplanar \ion{H}{1}
a few kpc away from the disk.  Finally, we emphasize that any galaxy with a
low luminosity ($L_X \lesssim 10^{39}$\,erg\,s$^{-1}$) X-ray halo (including
many nearby massive spirals) must have a low cooling rate of 
$\dot{M} < 1 M_{\odot}$\,yr$^{-1}$, with $\dot{M} < 0.1 M_{\odot}$\,yr$^{-1}$
for a temperature like that of the Galactic halo.  

Even if the fountain model can be reconciled with NGC\,6946
\citep{boomsma08}, NGC\,891 has a \ion{H}{1} halo four times more
massive with a similar instantaneous SFR.  The mass loading would then
need to be a factor of $4-8$ in order to balance the fallback rate of
the extraplanar \ion{H}{1}.  If the \ion{H}{1} within 1\,kpc of the
disk (but outside a few scale heights) is also participating in the
fountain, the mass loading factor must exceed about 10 (this may be
material that was not ejected with sufficient velocity to exceed
1\,kpc).  It is difficult to envision any fountain scenario in which
this could be the case.

Since cold accretion does not appear to be a viable source of
\ion{H}{1}, we speculate that a prior outflow is the best explanation.
Unfortunately, there are few constraints on the characteristics of
such an outflow, so it is an unsatisfactory, if plausible, explanation
at present.  However, if the present activity in the disk is
insufficient to produce the neutral halo, it is the best candidate.
Finally, if it is conclusively demonstrated that the X-ray emitting
gas is indeed low metallicity, a prior outflow scheme would decouple
it from the observed \ion{H}{1} halo (that presumably has the same
metallicity of the disk).

\section{Summary \& Conclusions}

We have used deep X-ray exposures of the nearby, edge-on spiral galaxy
NGC\,891 to determine the properties of its hot halo.  Our main
findings are summarized here:
\begin{itemize}
\item The hot halo is fit well by a thermal model with subsolar metallicity
 ($Z \sim 0.1 Z_{\odot}$) to high confidence.  Fits to the halo emission
 near the disk and farther out find nearly identical parameters, indicating
 a homogenous halo.  
\item The simplest explanation for a low-metallicity X-ray halo is hot 
 accretion from the IGM.
\item The best-fit values agree between the MOS, pn, and
 \textit{Chandra} ACIS-S detectors.  An examination of the X-ray
 background in the region indicates that it is properly subtracted
 from the spectra, and known systematic effects from model choices and
 fitting method do not substantially change our results.
\item A 2-T model with solar metallicity is also a good fit to the halo 
 spectrum when the emission measure is dominated by an absorbed, 
 cool, $kT \sim 0.1$\,keV component.  Most of the emission, however, comes 
 from a weaker $kT \sim 0.25$\,keV component.  
 The metallicity in this model supports a disk origin for the coronal
 material. 
\item When considering the X-ray cooling rate and \ion{O}{6} limits on the
 cooling mass as well as the density scale heights in the 1-T and 2-T
 models, the 1-T model is favored.
\item The aforementioned cooling rates are not high enough to offset the
 losses of the massive \ion{H}{1} halo due to fallback of cold material
 onto the disk.  This suggests that the extraplanar gas cannot be in
 equilibrium. 
\item An examination of single superbubble breakouts indicates that a
 galactic fountain cannot lift enough material into the halo to account
 for the massive \ion{H}{1} halo.   This corroborates the cooling rate
 problem.
\item There is no plausible cold accretion mechanism to supply so much
 gas so rapidly to NGC\,891, so we speculate that a prior outflow has
 taken place.  There is circumstantial evidence for some kind of 
 past nuclear starburst in a few other massive, late-type galaxies with
 high star formation spread throughout the disk.  However, the parameters
 of such an outflow cannot be constrained from the present data.
\end{itemize}

1-T and 2-T models of hot halos have been criticized as undermodeling
the true spectrum, which may be considerably more complex
\citep{strickland00b}.  However, there is reason to believe (at least
in non-starburst galaxies) that the halo really is simple.  Although
we have presented arguments in favor of the reality of the low $Z$ in
the 1-T model and that the 1-T model better explains the coronal
morphology, a definitive test could be made with a deep (400\,ks
total) exposure that would allow line ratios to be measured even at
the low spectral resolution of the EPIC CCDs.  We believe that we have
demonstrated that the background is well enough understood and the
temperature well enough established to make this test feasible.

In addition, such an exposure would allow a definitive study of the
smoothness of the halo.  Coronal material is frequently described as
``filamentary'' in the literature, but in many cases the appearance of
filaments may be biased by adaptive smoothing or related to PSF wings.
In our \textit{XMM-Newton} images of NGC\,891, some apparent
filamentary structure is revealed to be slightly brighter regions of
the halo connected by dim point sources (seen in the \textit{Chandra}
data) or arcs that clearly trace the edge of a subtracted point source
mask.  Since the presence of this filamentary structure is an
important test of state-of-the-art cooling models for accreted or
hydrostatic halos \citep[e.g.][]{mccourt11,hobbs12}, a deep search for
filaments that are not obviously related to outbursts (as in starburst
galaxies) would be a welcome contribution.

Progress on the steady-state front requires a more sophisticated
treatment of the galactic fountain mechanism, allowing for the action
of interacting superbubble breakouts and superbubbles that form
beneath debris falling back from prior breakouts.  If it still appears
that the instantaneous SFR in NGC\,891 and its high SFR counterparts
cannot explain the amount of extraplanar gas, a serious search for a
sequence of post-starburst galaxies with luminous X-ray halos is
warranted.  Improved \ion{H}{1} imaging from the HALOGAS project
\citep{heald11} will refine the estimates of \ion{H}{1} residing above
1\,kpc in nearby galaxies, but unfortunately X-ray imaging remains
sparse.  Still, the work by \citet{tyler04} and \citet{kuntz10}
indicates that most galaxies will not have X-ray halos nearly as
luminous as NGC\,891.  This in turn places cooling rate limits of about
$1 M_{\odot}$\,yr$^{-1}$ on the halos of these galaxies, assuming a
single-temperature plasma in hydrostatic equilibrium.

\acknowledgments

The authors thank the referee for his or her careful attention to this
work and for helpful suggestions that improved the discussion of the
\ion{H}{1} halo.  EHK gratefully acknowledges support from NASA ADAP grant \#061951.

{\it Facilities:} \facility{XMM} \facility{CXO}


\end{document}